\documentclass[iop]{emulateapj-rtx4} 
\shortauthors{Sekanina}
\shorttitle{Great Comets of 1843 and 1882 at Their Previous Return}
\slugcomment{Version \today }

\newcommand{\lapeq}{$\;$\raisebox{0.3ex}{$<$}\hspace{-0.28cm}\raisebox{-0.75ex}{$\sim$}$\;$}

\begin{document}
%
%
\title{The Great Comets of 1843 and 1882 at Their Previous Return to
 Perihelion\\in the Twelfth Century:\ One Spectacular, the Other
 Dull \\[-1.55cm]}
\author{Zdenek Sekanina}
\affil{La Canada Flintridge, California 91011, U.S.A.; {\sl ZdenSek@gmail.com}}
\begin{abstract} 
New insights into the history of C/1843 D1 and C/1882 R1, the two celebrated
Kreutz sungrazers, are provided by assessing evidence on their appearance at
the previous perihelion return, known as X/1106~C1 and the Chinese comet
of 1138 (Ho's No.~403), respectively.  The conditions differed vastly because
of disparities in geocentric distance, solar elongation, and phase correction
(forward~scattering), all linked to the arrival times (early February vs
early August).  The conclusions include:\ the daytime sighting of the 1106
comet by Sigebert de Gembloux is consistent with expectation and so are the
accounts of an exceptionally long tail observed later in twilight; the comet
reached perihelion only hours before its daytime detection; the 1138 comet
could have never been sighted in daylight~or~discovered much earlier than it
actually was, in early September, one month after perihelion; at discovery,~the
tail is predicted to have reached elevations of 15$^\circ$--30$^\circ\!$, while
the~head was~only~10$^\circ$ above horizon,~when observed from moderate
northern latitudes; the notion that Kreutz sungrazers at perihelion between
mid-May and mid-August could not be seen from the ground except possibly
in daylight~is~misleading; the appearance of the 1138 comet's nucleus after
its tidal fragmentation at perihelion is modeled on the assumption that it
consisted of five major fragments (including C/1882~R1 and C/1965~S1); and
unpredictable morphological changes with time in the 1882 sungrazer's split
nucleus are discussed.
\end{abstract}
\keywords{comets:\ X/1106\,C1, 1138, X/1702\,D1, C/1882\,R1, C/1965\,S1;
methods:\ data analysis{\vspace{-0.16cm}}}

\section{Introduction} 
Even though the perception of the Kreutz sungrazer system has over the past
decades been changing profoundly, one attribute has remained untouched
for nearly sixty years, since Marsden's (1967) pioneering work:\ regardless
of the number of populations (or subgroups) that the system is believed
to consist of, the two preeminent ones, incorporating most observed
members, have persisted.  Marsden called them Subgroups~I and II, and
they were the only ones known in the 1960s.  Significantly, of the two
most memorable sungrazers, both appearing in the 19th century less than
40~years apart, one was a member of Subgroup~I (now also called
Population~I) and the other a member of Subgroup~II (or Population~II).
I do of course refer to the {\it Great March Comet of 1843\/} (C/1843~D1)
and the {\it Great September Comet of 1882\/} (C/1882~R1).

The remarkable orbital work by Hubbard (1851, 1852) demonstrated that the
orbital period of the 1843 sungrazer was confined to a likely range of
500--800~years.  And the monumental effort by Kreutz (1888, 1891)
made it clear that the orbital periods of the four nuclei, the most
prominent products of the 1882~sungrazer's perihelion fragmentation,
ranged between 650 and 950~years.  When comet Ikeya-Seki (C/1965~S1),
another member of Population~II, arrived in the practically identical
orbit 83~years following the 1882~sungrazer and Marsden (1967)
determined its barycentric orbital period to equal about 850~years,
it was likely that (i)~the two objects had separated from one another
early in the 12th century and (ii)~the orbital period of nucleus~B
(or No.~2 in Kreutz's notation) of the 1882~sungrazer, of about 770~years,
almost coincided with the orbital period of the comet before it split at
perihelion.  The union of the 1882~sungrazer and Ikeya-Seki in a single
body prior to their 12th-century perihelion was proven by Marsden
rigorously, via direct orbit integration; nucleus~B was obviously
the largest mass of the 1882~sungrazer.{\vspace{0.05cm}}

Given that sungrazers have been appearing over many centuries, even
millennia, the gap of less than 40~years between the perihelion times of
the presumably most massive members of the Kreutz system looks suspiciously
small to say the least.  This near coincidence makes one speculate that
--- in analogy to the Population~II pair --- the gap could be
an accumulated effect of the rate of their separation in the past.
However, compared to the 1882~sungrazer and Ikeya-Seki, two major
discrepancies become apparent:~(i)~the small gap accumulated over
the time shows that the orbital periods of the 1843 and 1882 comets
were very similar, whereas (ii)~their other orbital elements differed
much more.  These arguments~prompt one to conclude that the putative
breakup of the progenitor comet into its two massive survivors was
characterized by an orbital-momentum exchange with a significant
contribution in the direction normal to the orbital plane and an
orbital-plane contribution that affected appreciably the perihelion
distance but only marginally the orbital period.  It turns out that
these conditions are satisfied only if the fragmentation event took
place at large heliocentric distance.{\vspace{0.05cm}}

The similarity of the orbital periods suggested that the 1843 and 1882
sungrazers had passed their previous perihelion, apparently in the
early 12th century, in the same order and at times even closer to one
another than in the 19th century.  Since only one suspect in the
early 12th century had until recently been known --- the
Great Comet of 1106 (X/1106~C1) --- we were confronted
with the problem of a {\it missing second sungrazer\/}.

\section{Search for the Missing Sungrazer:\\Past Investigations} 
An important orbital property common to all Kreutz sungrazers is
their fixed direction of the line of apsides, except for a minor
difference of about 0$^\circ\!$.5 between Populations~I and II.
For example, the line of apsides of the Great March Comet of 1843
is defined by the ecliptic longitude and latitude of perihelion
equaling, respectively, 282$^\circ\!$.58 and +35$^\circ\!$.29
(equinox J2000), showing that the comet arrived from the direction
of the constellation of Canis Major, a point a few degrees north
of Sirius.

It is well known that, in general, the southern~hemisphere is
favored and that the observed performance of a sungrazer of the
given intrinsic brightness depends strongly on the angle that
the line of apsides makes~with the Sun-Earth line at the time
the comet is at perihelion, implying significant seasonal
variations.~\mbox{Because} essentially all preperihelion ejecta are
sublimated away at close proximity of perihelion, sungrazers
become spectacular objects along the post-perihelion
branch of the orbit because of their extremely high rates of
dust emission primarily in the first hours after perihelion.
An intrinsically bright sungrazer makes a spectacle in the
evening sky when its perihelion occurs in January--March, in
the morning sky when in September--November.  December arrivals
reach the southernmost areas of the sky.

Problematic is the appearance of Kreutz sungrazers passing perihelion
at ``wrong'' times of the year. While I do not fully subscribe
to it, the verdict found in most cometary publications is closely
matched by Marsden's (1967) narrative that a sungrazer ``{\it at
perihelion between mid-May and mid-August will undoubtedly be
missed, unless it can be seen in daylight.  The geometry is such
that the comet would approach and leave the sun from behind, and
it would never be seen in a dark sky\/}.''

One of the issues that I am having with this widely~accepted
notion involves the comet of 1041 in Hasegawa \& Nakano's (2001)
list of suspected sungrazers.  This object, observed in China
and Korea from September~1 on, was estimated by the authors
to have passed perihelion on August~4\,$\pm$\,4, more than a
week before the end of the disadvantaged time zone.  No daylight
sightings were reported, but the comet {\it was\/} observed about
four weeks after perihelion, when Hasegawa \& Nakano computed its
position to have been some 1.7~AU from the Earth, 1~AU from
the Sun, and a solar elongation of 35$^\circ$.

When Sekanina \& Kracht's (2022) in-depth investigation of the motion
of Ikeya-Seki (C/1965~S1) showed that the comet previously was at
perihelion around the year 1140 and the Great Comet of 1106 (X/1106~C1)
could not be the parent of either Ikeya-Seki or the Great September Comet
of 1882, perfunctory search in Ho's (1962) catalogue resulted in the
discovery of a Chinese comet of 1138, under No.~403, as the likely
candidate.  Entirely by coincidence, its timeline strongly resembled
the timeline of the comet of 1041:\ the first observation on September~3
and an estimated perihelion time on August~1.

Detailed examination of the conditions for the likely obsevation
site indicated that at midpoint of the astronomical twilight,
about 4:30 local time on 1138 September~3, the comet's head
would have been located, as reported, in the east.  Its predicted
elevation was about 11$^\circ$, its solar elongation 36$^\circ$,
and it was nearly 1~mag brighter than a computed
limiting magnitude of 3.2 for the naked eye; the brightest part
of the dust tail was expected to be approximately 8$^\circ$ long at
position angle near 270$^\circ$, pointing up at an angle with the
vertical, its end at least 15$^\circ$ above the horizon.  To the
naked eye it was a rather unimpressive object, even though the
adopted intrinsic brightness on its approach to perihelion equaled
that of comet X/1106~C1.  There was no report of daylight sighting,
again paralleling the case of the 1041 comet.

Reports of the detections of these two objects in a dark sky (or
nearly dark sky of the astronomical twilight) and no daylight
detection reports appear to contradict the above generally accepted
viewpoint of the Kreutz sungrazers at perihelion in mid-May through
mid-August, and this controversy needs to be examined.

Although the comet of 1041 is of major interest as a possible previous
appearance of comet Pereyra (Sekanina \& Kracht 2022), the primary
subject of this paper~is~the Chinese comet of 1138 as the previous
appearance of the Great September Comet of 1882 and comet Ikeya-Seki.
In the following I compare the 1138 comet with the Great Comet of
1106; this is particularly appropriate because both objects are
believed to have been of about equal intrinsic brightness, yet the
1106 comet was seen in broad daylight $\sim$1$^\circ$ from the Sun on
February~2 and displayed a spectacular tail up to $\sim$100$^\circ$ long
in the following weeks (e.g., Kronk 1999), while the 1138 comet was
a lackluster object, reported for the first time more than 30~days
after the estimated perihelion time.  Yet, the 1106 comet disappeared
some 45--50~days after perihelion, while the 1138 comet about 60~days
after perihelion.  This effect may have been a by-product of the 1138
comet's more extensive perihelion fragmentation, which is believed to
lead to a slower rate of post-perihelion fading (Sekanina 2022;
Section~3.1).

\section{Comets of 1106 and 1138 Shortly After Passing\\Perihelion, and
 the Limiting Magnitude} 
The part of a bright comet that has the greatest chance to be seen
with the naked eye in broad daylight is the head because of its
high surface brightness.  In twilight and at night, the most obvious
part is the tail because of its extent and greater angular distance
from the Sun in the sky.  To compare the positions, relative to the
Sun and Earth, of the heads and tails of the two comets, I begin with
their adopted orbital elements, presented in Table~1.  The set for X/1106~C1
comes from Sekanina \& Kracht's (2022) integration of the motion of the
Great March Comet of 1843; for the Chinese comet of 1138 from their
integration of the motion of nucleus~B of the Great September Comet of
1882.  The perihelion time of the 1106 comet, derived by Hasegawa \&
Nakano (2001), was adjusted to better fit the early observations.

Next, I derive the conditions under which observations of the two
comets' heads were, or would have been, made over the first ten
days after perihelion.  For either comet, the columns in Table~2
provide at each time the heliocentric and geocentric distances,
apparent position relative to the Sun in the sky in{\vspace{-0.04cm}}
the polar coordinates (solar elongation and position angle), and phase
angle.\footnote{Even though an adopted orbit for the 1106 comet is
presented in Table~1 and used in Table~2, the object is still
designated with X/, since no orbit derived from the 1106
observations is available.\\}

\begin{table}[t] 
\vspace{0.12cm}
\hspace{-0.19cm}
\centerline{
\scalebox{1}{
\includegraphics{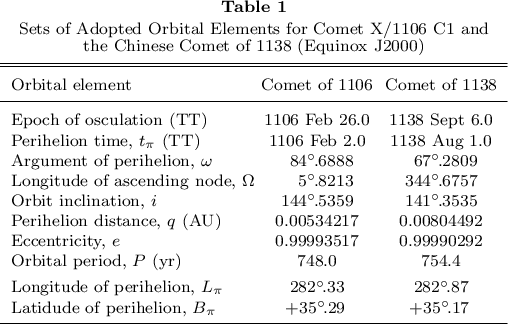}}}
\vspace{0.55cm}
\end{table}

Table 2 demonstrates that the unfavorable geometry~of the 1138 comet
affected its observing conditions in three different manners:\ the
comet was getting fainter as the geocentric distance kept increasing,
but its light was also dimmed by backscattering.  In addition, as the
rate of angular separation from the Sun was slowing down by the growing
distance from the Earth, the comet was projecting on a relatively
brighter sky background.  Conversely, the favorable geometry helped
the 1106 comet look brighter not only because of the decreasing
geocentric distance (the first seven days), but also on account of
forward scattering, especially in the early hours after perihelion.
And because the comet was receding fairly rapidly from the Sun in
the sky, the period of daylight appearance was cut short.

In the following, the observing conditions of the 1106 and 1138
comets are examined quantitatively.  To establish whether the
star-like head of a sungrazer could be detected in broad daylight
with the naked eye, I employ Schaefer's (1993, 1998) algorithm
for the visual limiting magnitude of stellar objects.  Its
application requires that certain approximations be introduced
first.

\subsection{Adopted Light Curves of the 1106 and 1138 Comets} 
A procedure for generating approximate light curves of historical
sungrazers or their candidates, introduced~in Sekanina (2022),
was based on a limited amount of information available on the
light curves of a few bright sungrazers, mainly Ikeya-Seki and
the Great September Comet of 1882, to a lesser extent also the
Great March Comet of 1843.  Broadly, the formula was in line
with the well-known fact that the Kreutz sungrazers were fading
rapidly after perihelion unless they split at perihelion:\ the
rate of fading was assumed to correlate with the number of
persisting nuclear fragments seen after perihelion.

\begin{table}[t] 
\vspace{0.12cm}
\hspace{-0.19cm}
\centerline{
\scalebox{1}{
\includegraphics{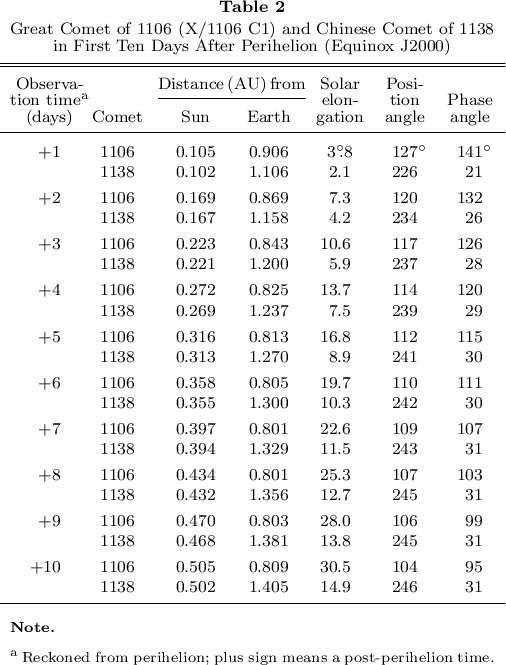}}}
\vspace{0.5cm}
\end{table}

Applied below is a slightly expanded version, which includes a phase
correction.  At a zero phase angle, when the correction by definition
vanishes, the light curves of the 1106 and 1138 comets were assumed to
comply with two postulates:\ (i)~they were identical before perihelion,
predicated on the premise that the nuclei of the 1843 and 1882
sungrazers comprised, respectively, about 40~percent of each parent's
mass; and (ii)~the 1138 comet, a Population~II member, split at
perihelion into more persisting fragments than the 1106 comet, a
Population~I member.  Because of the fragmentation, the light
curves were not, in general, symmetric relative to perihelion.

Given that the absolute preperihelion magnitudes of the Great March
Comet of 1843 and the Great September Comet of 1882, normalized to
1~AU from both the Sun and Earth, $H_0^-$, were adopted in Sekanina
(2022) to equal 3.5 and 3.4, respectively, the absolute preperihelion
brightness of the parents{\vspace{-0.01cm}} should be greater in
proportion to the surface area,{\vspace{-0.1cm}} or mass to the power
of $\frac{2}{3}$, that is \mbox{$(1/0.4)^{\frac{2}{3}} \!=\! 1.84$},
or by 0.66 mag.  Averaging, I get~for~both the 1106 and 1138 comets
\mbox{$H_0^- \!= 3.45 - 0.66 \simeq 2.8$}.~In~line with the observed
light curve of Ikeya-Seki, the preperihelion brightness variations
with{\vspace{-0.085cm}} heliocentric distance $r$ were assumed to
follow an $r^{-n^-\!}$ law, where \mbox{$n^- \!= 4$}.  The apparent
preperihelion{\vspace{-0.02cm}} magnitude, $H_{\rm app}^-(r,\Delta,\alpha)$,
for either of the two 12th century comets was given by the expression
\begin{equation}
H_{\rm app}^-(r,\Delta,\alpha) = 2.8 + 10 \log r + 5 \log \Delta
 + \Phi(\alpha),
\end{equation}
where $\Delta$ is the geocentric distance, $\alpha$ is the phase
angle, and $\Phi(\alpha)$ is the phase correction, computed from the
standard law for dust-rich comets, formulated by~\mbox{Marcus} (2007).
Both $r$ and $\Delta$ are in AU and \mbox{$\Phi(0^\circ) = 0$}.  It
certainly is possible that the light curve flattens and the law in
Equation~(1) overestimates the actual brightness at extremely small
heliocentric distances.  However, sungrazers spend very little time,
always less than one day, at close proximity of the Sun, so that
this approximation is usually not a problem in practice.

After perihelion{\vspace{-0.07cm}} the intrinsic brightness is assumed
to vary with heliocentric distance as $r^{-n^+}$, where in general
\mbox{$n^+ \!\neq n^-$}.  The post-perihelion light curve must
connect with the preperihelion light curve at perihelion, and the
condition for the absolute post-perihelion magnitude, $H_0^+$, reads
\begin{equation}
H_0^+ \! = H_0^- \! + (10 - 2.5 \, n^+) \log q,
\end{equation}
where $q$ is the perihelion distance in AU.  Along the post-perihelion
branch of the orbit, the slope $n^+$ of the light curve is assumed to
be a function of nuclear fragmentation at perihelion, decreasing with
the increasing number $\nu_{\rm frg}$ of persisting fragments.  A
linear law, fitting the data on (a)~comet Ikeya-Seki, which displayed
two nuclei after perihelion and whose \mbox{$n^+ \!= 4.0$}, and (b)~the
1882 sungrazer with five to six fragments and \mbox{$n^+ \!= 3.3$}
(Sekanina 2002), is given by
\begin{equation}
n^+ \!= 4.4 - 0.2 \, \nu_{\rm frg}.
\end{equation}

For objects that do not split at perihelion, such as the 1843 comet,
\mbox{$\nu_{\rm frg} = 1$}, and \mbox{$n^+ \!= 4.2$}.  This is in line
with bright sungrazers' fading more rapid than their brightening in
the absence of perihelion splitting.  However, this procedure is
inapplicable to the dwarf sungrazers, such as those seen in images
taken with the coronagraphs on board the Solar and Heliospheric
Observatory.

The possible perihelion-fragmentation histories of the 1106 and 1138
comets were addressed in one of my recent papers (Sekanina 2025).
I speculated that besides the Great March Comet of 1843, the main
returning mass of the Great Comet of 1106, comet C/1668~E1 was a
major preceding fragment and another large piece could arrive at
perihelion around the year 2050.  Their possible early sibling ---
a comet of 1529, and another one, which might arrive in the
mid-23rd century, are just two rather remote possibilities.
Adopting \mbox{$\nu_{\rm frg} = 3+$}, I get \mbox{$2.5\,n^+ \! \simeq
9.3$} from Equation~(3).  With the perihelion distance from
Table~1, the absolute post-perihelion magnitude of the
1106 comet is \mbox{$H_0^+ \!= +1.2$} from Equation~(2).

In the same recent paper I proposed that besides the Great September
Comet of 1882 and comet Ikeya-Seki, another likely fragment of the
Chinese comet of 1138 was X/1702~D1 and one to return to perihelion
before 2040, possibly even before the end of this decade.  Uncertain
as a related piece was a ``sun-comet'' recorded in China's Shandong
Province in April or May, 1792 (Strom 2002).  I adopt \mbox{$\nu_{\rm
frg} = 4+$} and \mbox{$2.5\,n^+ \! \simeq 8.9$}.  For{\vspace{-0.06cm}}
the absolute magnitude I get \mbox{$H_0^+ \! = +0.5$}.  Accordingly,
the post-perihelion light curves become
\begin{eqnarray}
H_{\rm app}^+\! & = & 1.2 \!+\! 9.3 \log r \!+\! 5 \log \Delta \!+\!
 \Phi(\alpha) \;\,\ldots\;\, {\rm for\;X/1106\:\,C1}, \nonumber \\[-0.11cm]
H_{\rm app}^+\! & = & 0.5 \!+\! 8.9 \log r \!+\! 5 \log \Delta \!+\!
 \Phi(\alpha) \;\,\ldots\;\, {\rm for\;1138\;comet}. \nonumber \\[-0.07cm]
  & & 
\end{eqnarray}
The assumptions of an identical preperihelion light curve with the 1106
comet and a stronger propensity for fragmentation at perihelion leads to
the 1138 comet being intrinsically brighter after perihelion, by nearly
1~magnitude at 1~AU from the Sun.

\begin{figure}[t] 
\vspace{0.16cm}
\hspace{-0.2cm}
\centerline{
\scalebox{0.65}{
\includegraphics{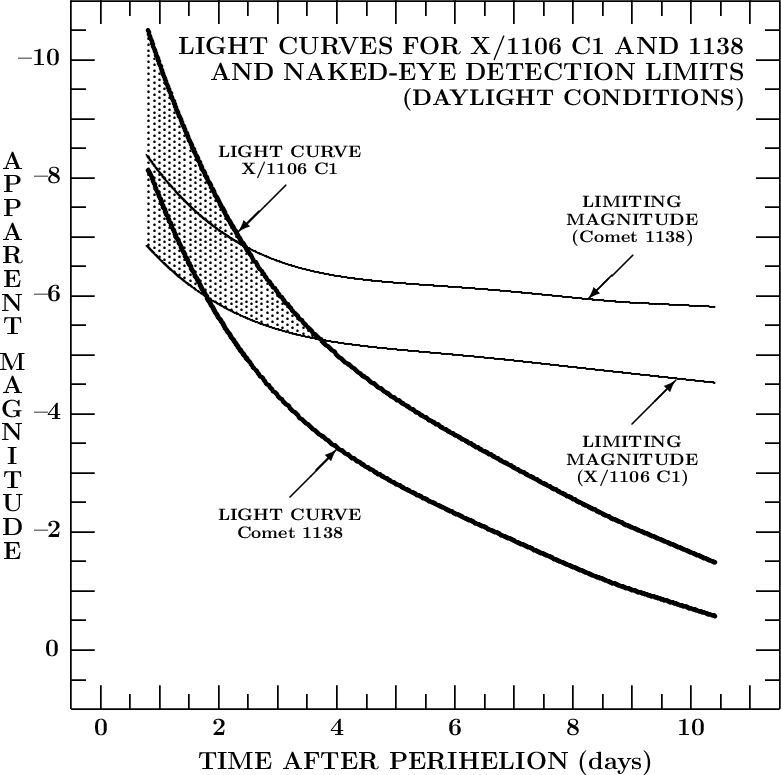}}}
\vspace{0cm}
\caption{Conditions for daylight observations of the 1106 and 1138
comets in the first 10~days after perihelion.  Even though
the 1138 comet was intrinsically brighter after perihelion, the
greatly superior observing conditions made the 1106 comet~sub\-stantially
brighter in daylight, as shown by comparing the light~curves (thick
lines) with the limiting magnitudes for the naked eye (thin lines) at
local noon.  The shaded area predicts that the favorable daylight
conditions for detecting the 1106 comet may have prevailed over the first
three days after perihelion.  The 1138 comet is predicted to have been
undetectable with the naked eye in daylight.{\vspace{0.55cm}}}
\end{figure}

\subsection{Site Data For Limiting-Magnitude Prediction} 
In Schaefer's algorithm, the visual limiting magnitude is a function of
four parameters that depend on the observing site and the conditions
at the time.  For the exercise I choose the only location where both
comets were observed, which was the territory of the Sung Dynasty in
China.  The geographic data for the putative observing{\nopagebreak}
site should adequately be approximated by an average for two modern
Chinese cities:\ Nanjing, the capital of Jiangsu Province, and Hangzhou,
the capital of Zhejiang Province.

The data needed for the algorithm are the geographic latitude, for
which I use +31$^\circ$; the altitude above sea level of 20~meters.
For daylight conditions I adopt the air temperature and relative
humidity of, respectively, 9$^\circ$C and 74\% in early February
for the 1106 comet and 31$^\circ$C and 79\% in early August for
the 1138 comet.

\subsection{Visibility of the Comets in Broad Daylight} 
The effects of the Moon in broad daylight were ignored, because when
near the comet in the sky, the Moon must have been near new and its
contribution negligible.  The Moon was new on 1106 February~5.95~UT
and on 1138 August~7.62 and September~6.29~UT.  Except near the horizon,
the limiting magnitude depends only weakly on the Sun's and comet's
zenith distances, and I considered the case of both the Sun and the
comet near the local meridian and their zenith distances equal.
For such conditions, the implied zenith distances were
$\sim$50$^\circ$ for the 1106 comet and $\sim$10$^\circ$ for the
1138 comet.

The apparent light curves of the two comets and the respective limiting
magnitudes are plotted in Figure~1.  The striking feature is the
reversed order of the light curves.  Whereas Equations~(4) showed the
1138 comet to be almost 1~mag {\it intrinsically\/} brighter at 1~AU
after perihelion, Figure~1 predicts that the 1106 comet should~have
{\it appeared\/} distinctly brighter to the terrestrial observer.
The driver of this reversal was the differential phase effect early
after perihelion, while later the disparity in the geocentric
distance became more important.  And, as follows from Table~2, for
a given heliocentric distance (which governs the intrinsic-brightness
variations) the solar elongation of the 1138 comet was always only
about one half of the solar elongation of the 1106 comet, a major
factor affecting the visibility of the two objects.

An important point to make is that the unfavorable geometry appears
to affect the visibility of a sungrazer {\it most severely in close
proximity of the Sun\/}.  Indeed, while the 1106 comet was brighter
than the limit for naked-eye detection over the first three days
after perihelion, the 1138 comet stayed {\it in daylight\/} below
this threshold at all times.  The experience with the 1138 comet,
including its comparison with the 1106 comet, thus suggests that
the widely-accepted perception that a sungrazer at perihelion between
mid-May and mid-August
{\it will be missed, unless it can be seen in daylight\/}
is misleading and should be {\bf replaced} with
{\bf will be missed in daylight}
and the rest of the statement, namely,
{\it will never be seen in a dark sky\/}
should be {\bf deleted}.  These comments fundamentally change the
status of the sungrazers passing through perihelion between mid-May
and mid-August.

\begin{figure}[b] 
\vspace{0.8cm}
\hspace{-0.2cm}
\centerline{
\scalebox{0.64}{
\includegraphics{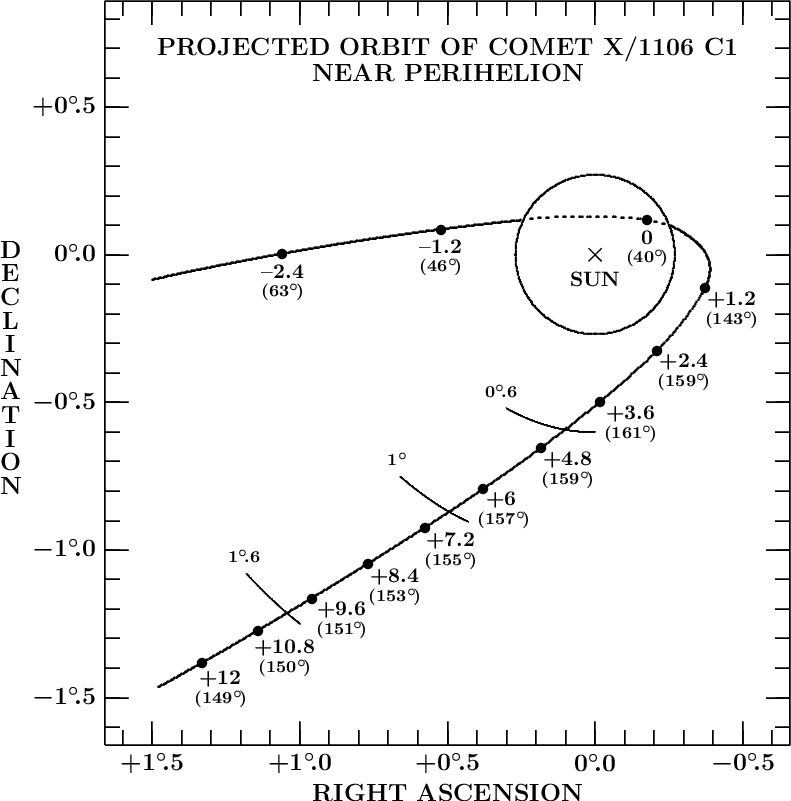}}}
\vspace{0cm}
\caption{Projection onto the plane of the sky of the predicted orbit for
X/1106 C1 close to perihelion (equinox J2000).  The dotted part of the orbit
indicates that the comet was then behind the Sun.  The bullets show the
locations of the comet at the times,  reckoned in hours from perihelion.
Parenthesized are the respective phase angles.  Also shown are arcs of
solar elongations of~0$^\circ\!\!$.6,~1$^\circ\!$,~and 1$^\circ\!\!$.6.
Receding from the Sun at a nearly constant projected rate of $\sim$4$^\circ$
per day starting shortly after perihelion, the comet was observed at
Gembloux in daylight as a star over a period of six hours.{\vspace{0.1cm}}}
\end{figure}
%

%
\section{Daytime Observation of X/1106 C1\\at Gembloux, Belgium} 
The extremely favorable daylight observing conditions predicted for
the 1106 comet appear to have facilitated an amazing sighting, described
by Sigebert de Gembloux in his {\it Chronicon sive Chronographia\/} in
AD~1111.~A~monk~in the Benedictine Abbey of Gembloux, Belgium, he
noted in the chronicle that {\it on 1106 February~2 a star appeared
during the daytime, between the third and ninth hour, about a cubit
from the Sun\/}.\footnote{To answer the question of what was the distance
in degrees is not straightforward.  One can find a lot of information
on the {\it cubit\/} in the literature, including the differences
between the Sumerian cubit, Egyptian royal cubit, Greek cubit, Roman
cubit, etc.\ and anthropologists still write scientific papers on the
cubit in the 21st century (e.g., Stone 2014).  Yet, all this information
is irrelevant to Sigebert's observation, because the primary usage of
the term cubit in the antiquity was as a {\it unit of length\/},
equal to the extent of the forearm from the tip of the middle finger
to the elbow.  Different cubits had different lengths, mostly between
17.5 and 21 inches, or 44 and 53~cm. --- Obviously, the cubit as a
measure of length and the cubit as a measure of angular separation are
two very different quantities, and it is unfortunate that for some
strange reason the same term is used for both of them.  It is more
difficult to find information about the cubit measuring angular separation.
And this information is not only scarce but, worse, also inconsistent.
When trying to google the term, one usually finds that 1~cubit is
approximately 1$^\circ$, with no details or reference being given.
There are, however, exceptions.  To an inputed request ``cubit angular
measure'' Google's AI Overview returns:\ {\it A ``cubit'' as an angular
measure roughly corresponded to a finger's width held at arm's-length,
which subtends an angle of about 2~degrees.}  Similarly, in Wikipedia's
article on ``Cubit'' one can read:\ {\it There is some evidence that
cubits were used to measure angular separation. The Babylonian
Astronomical Diary for 568--567~BCE refers to Jupiter being one cubit
behind the elbow of Sagittarius. One cubit measures about 2~degrees\/},
with a reference to Steele (2008), noting that the relationship between
the cubit as a unit of length and a unit of angular separation was not
in the book elaborated on.  Interestingly, I was unable to find the
cubit, as an angular measure, in Wikipedia's article ``List of obsolete
units of measurement,'' or among the angular measure units in Wikipedia's
article ``List of unusual units of measurement,'' or in any of the
associated articles ``Ancient Arabic (Egyptian, Greek, Mesopotamian,
Roman) units of measurement.'' (I have not investigated sources
concerning the works by Otto E.\ Neugebauer, as this would involve
prohibitive time demands.)  On the other hand, in their paper on ancient
Babylonian time measurements Stephenson \& Fatoohi (2008) wrote
that it {\it should be noted that observed angular distances
between celestial bodies were almost invariably expressed by the
Babylonians in terms of the k\`{u}\v{s} (cubit), which was equivalent
to between about 2 and 2.5 degrees\/}.  The various sources clearly
provide a rather wide range of numbers.  Using the defining ratio of
the finger's width to arm's-length,{\vspace{-0.04cm}} taking the
standard values of 1.85~cm for the finger's width{\vspace{-0.05cm}}
and $\frac{14}{9}$ for the ratio of arm's-length to forearm's-length
of 46~cm, I derive that \mbox{1~cubit = 1$^\circ\!$.5}, yet another
estimate.}

Although the comet was called a star by Sigebert, a tail up to 100$^\circ$
long several days later left no one doubting the nature of the object
(Section~5).  It is common that in broad daylight only a very short
or no tail is seen with the naked eye because of the low surface
brightness.~What is remarkable about the statement is the time
constraint.  Sigebert clearly reckoned the time from sunrise.  He
said nothing about the first three hours, but that means nothing;
the Sun may have been shrouded in clouds or fog, or the separation
was too small for optical discrimination in the Sun's glare, or it
simply did not occur to him (and/or others) to look up the sky earlier.

In any case, to inspect the timeline more closely, I plot in Figure~2
the orbit of the 1106 comet in projection onto the plane of the sky from
3.5~hours before perihelion to nearly 13~hours after perihelion.  For
approximately one hour, the comet was hidden behind the Sun, including
its perihelion point.  When observed at the Abbey by Sigebert himself
and/or his colleagues, the comet was already past perihelion and moving
to the southeast of the Sun at a nearly constant rate of $\sim$10$^\prime$
per hour in the general direction of the Earth.

It turns out that from Sigebert's description one can approximately
establish the UT times of observation and therefore the comet's perihelion
time.  Given that the geographic longitude and latitude of the Abbey were,
respectively, 4$^\circ\!$.70\,E and 50$^\circ\!$.56\,N, and the elevation was
150~meters above sea level, the Sun rose at 7:13~UT and set at 16:42~UT on
1106 February~2 (not corrected~for refraction).  Daylight thus extended over
9:29~hours and, rounded to the nearest quarter of an hour, the {\it comet
was seen between 10:15 and 16:15~UT\/}.  Since the latter time was less than
30~minutes before sunset, it is reasonable to conclude that the event's
reported termination was a corollary to the deteriorating atmospheric
conditions at low elevations, not the comet's sudden fading.  During the
six-hour observing window, the comet's solar elongation, $\epsilon$, is
predicted from Figure~2 to have increased by slightly more than 1$^\circ$.
The relationship between $\epsilon$ and the time from perihelion, $\tau$
(in days), is given more accurately by an empirical formula:
\begin{eqnarray}
\epsilon & = & -0^\circ\!.1904 + 4^\circ\!.3822 \,\tau - 0^\circ\!.3351
 \:\! \tau^2 \!, \\[-0.1cm]
	 &   & \pm 0^\circ\!.0042 \,\pm\! 0^\circ\!.0185
 \;\;\;\:\! \pm\! 0^\circ\!.0186 \nonumber
\end{eqnarray}
with a mean error of $\pm$0$^\circ\!$.0044 or $\pm$0$^{\:\!\prime\!}$.26
and applicable to an interval of \mbox{+0.18 day\,$\leq \! \tau \!
\leq$\,+0.80 day}.  With the solar elongation of 0$^\circ\!$.6
about 4.4~hours after perihelion and 1$^\circ\!$.6 about 5.7~hours
later, the comet would have reached an elongation of 1~cubit equaling
1$^\circ\!$.1 some 7:15~hours after perihelion, equivalent to 13:15~UT,
the middle of Sigebert's six-hour long window.  One finds in this case
that the comet had passed perihelion at 6:00~UT, or on February~2.25~UT.

Similarly, if 1~cubit equaled 2$^\circ$, the comet would~have reached
the solar elongation of 1~cubit 12:30~hours after perihelion on
February~2.03~UT, about 5.5~hours earlier than in the previous case.
To summarize, the uncertainties in the source data (Sigebert's rounding
off the times to 1~hour; giving the solar elongation as {\it about\/}
1~cubit; the large error in the cubit-to-degree conversion; etc.) suggest
that the perihelion time has been established with estimated accuracy of
$\pm$0.3~day.

Although it was rather obvious from Figure 1 that just hours after
perihelion the 1106 comet's brightness should have greatly exceeded the
visibility threshold near local noon, it remained unanswered whether
this also applied at larger zenith distances.  Besides, Figure~1
was plotted for a site at a geographic latitude very different from
that of the Belgian site.  To address this problem I computed the
comet's apparent magnitude from the adopted light curve in Equation~(4)
as a function of time between sunrise and sunset on 1106 February~2
and then compared it with the limiting magnitude in Figure~3.  As the
site constants for the Abbey of Gembloux I employed the geographic
coordinates and elevation mentioned above, and adopted an average
temperature of +4$^\circ$C,\footnote{The limiting magnitude depends
only weakly on the temperature (with typical rates on the order
of 0.01~mag per $^{\!\circ}$C), so that a constant-temperature
approximation is, except near sunrise and sunset, deemed acceptable.}
and relative humidity of 89\%.  Over the period of 9.5~hours~between
sunrise and sunset the comet's brightness is predicted to
have dropped by $>$6~mag, its solar elongation to have increased by
almost 1$^\circ\!\!$.5, and its zenith distance to have stayed
above~66$^\circ$.

\begin{figure}[t] 
\vspace{0.17cm}
\hspace{-0.2cm}
\centerline{
\scalebox{0.585}{
\includegraphics{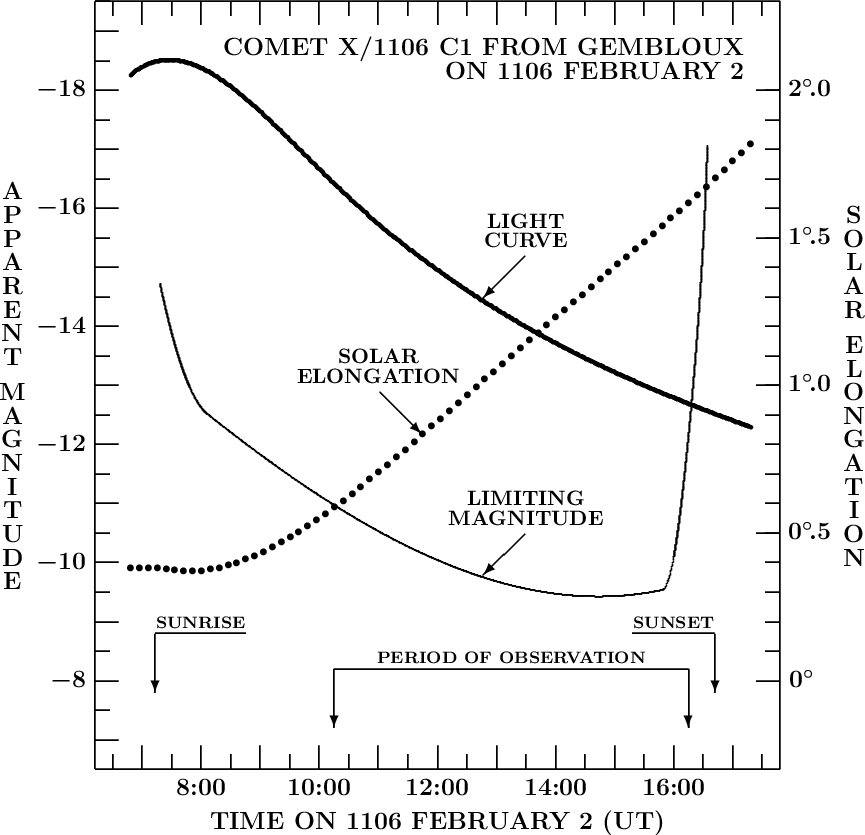}}}
\vspace{0cm}
\caption{Light curve (thick line), limiting magnitude (thin line),
 and solar elongation (dotted line) predicted for comet X/1106 C1
 vs time on 1106 February 2, when the comet was observed as a star
 over a period of six hours in broad daylight at close proximity of
 the Sun, according to a historical record by Sigebert, a monk in
 the Benedictine Abbey of Gembloux, Belgium.  The comet's brightness,
 greatly enhanced by the effect of forward scattering (up to nearly
 4~mag at maximum) is predicted to be some 4--6~magnitudes above the
 visibility threshold for the naked eye, except at times of very low
 elevation above the horizon, just after sunrise and before sunset.
 The start and end of the six-hour long period of observation
 are known with only limited accuracy.  The comet was assumed to pass
 perihelion at 6:00~UT on February~2.{\vspace{0.55cm}}}
\end{figure}

Figure 3 shows that the comet's brightness stayed above the predicted
visibility threshold by 4--6~magniudes except when very low above the
horizon.  However, it should be noted that the comet profited greatly
from effects of forward scattering, which at peak augmented its
brightness by nearly 4~magnitudes, about 3~hours after perihelion,
and contributed more than 2.5 magnitudes throughout the period of
time between 7:30 and 16:30~UT.  Interestingly, the end of Sigebert's
observation coincides with the time when the comet's light curve meets
the limiting magnitude.  Inaccuracies in Sigebert's report affect
choice of the perihelion time, which influences the positions of the
three curves in Figure~3.  For an earlier time, the comet's predicted
solar elongation would be greater and the comet would be fainter,~but
the visibility limit would drop as well.  In any case, the comet
must have offered an incredible sight and Sigebert, in retrospect, left
a valuable piece of scientific data.

An interesting question to which I do not have an answer is how much
of a problem for this interpretation~is Sigebert's silence about the
comet's increasing solar elongation over the six-hour period.  Is it
reasonable to expect that the gradual increase by up to 1$^\circ$
was large enough that it should have been detected, or is this {\it
merely\/} evidence that Sigebert's observations were not very careful?
Or could his silence favor a larger separation (i.e., more degrees
per Sigebert's cubit), because an increase in the separation by the
same amount is then harder to notice?

\section{Tails of the 1106 and 1138 Comets:\ Comparison} 
Under favorable conditions, a massive sungrazer passing perihelion
does not need more than just a few days to develop a tail, which 
becomes spectacular~over~a~short period of time and brighter than the
comet's head for weeks.  The main contributor in terms of the light
\mbox{effect}~is~dust, while plasma is present mainly early after
perihelion.  For the terrestrial observer to witness a prominent dust
tail, three conditions must be satisfied.  Any sungrazer complies with
the first condition:\ in order to produce a long tail, microscopic
dust grains that populate it must be rapidly transported to gigantic
distances from the comet's nuclues, the task accomplished by immensely
high accelerations furnished by solar radiation pressure at very small
heliocentric distances.  If the acceleration on a submicron-sized
particle{\vspace{0cm}} equals, say, 0.4~cm~s$^{-2}$~at~1~AU from the
Sun, it skyrockets to 40~m~s$^{-2}$ at 0.01~AU from the Sun.  The second
condition is satisfied only by massive sungrazers:\ sufficiently large
amounts of microscopic dust must be injected into the tail to make it
prominent, which a small fragment nucleus is incapable of.  And
the third condition requires that the tail point towards the
Earth:\ the effect perceived by the terrestrial observer becomes
then overwhelming, because the tail not only appears brighter
and its projected extent more impressive on account of its
smaller geocentric distance, but also because by virtue of its being
located between the Sun and Earth it is subjected to effects
of forward scattering, thereby further enhancing its brightness.
The 1106 and 1138 comets are nearly extreme examples that demonstrate
the utmost importance of the geometry.{\vspace{0.07cm}}

One prerequisite for examining the dust tails is knowledge of the
properties of the debris they contain.  Fortunately, the highly
relevant data --- on the tails of the Great March Comet of 1843, the
Great September Comet of 1882, and comet Ikeya-Seki of 1965 --- have
extensively been studied (e.g., Schwab 1883, Bennett \& Venter 1966,
Hagughney et al.\ 1967, Matyagin et al.\ 1968, Weinberg \& Beeson
1976a, 1976b, Krishna Swamy 1978, Warner 1980, Krishan \& Sivaraman
1982, Sekanina 2023).  Also investigated have been the dust tails
of a number of dwarf sungrazers (e.g., Sekanina 2000, Thompson 2009,
2015), imaged by the coronagraphs on board the Solar and Heliospheric
Observatory (SOHO); most of these objects appear to be debris of the
1106 comet (Sekanina 2024).\,\,\,{\vspace{0.07cm}}

The results on the sungrazers' dust tails are remarkably consistent in
that they show that the highest radiation pressure accelerations that
particles are subjected to are close to \mbox{$\beta = 0.6$} the solar
gravitational acceleration, compatible with the presence of
submicron-sized silicate grains, primarily olivine and pyroxene.
The other important finding is that all three major sungrazers
had a mixed tail, including both dust and plasma, at least early
after perihelion.  The plasma tail was then extending farther
than the dust tail, thus determining the feature's overall
length.  As the heliocentric distance grew, the plasma tail was
gradually disappearing.{\vspace{0.07cm}}

\begin{figure}[t] 
\vspace{0.17cm}
\hspace{-0.2cm}
\centerline{
\scalebox{0.72}{
\includegraphics{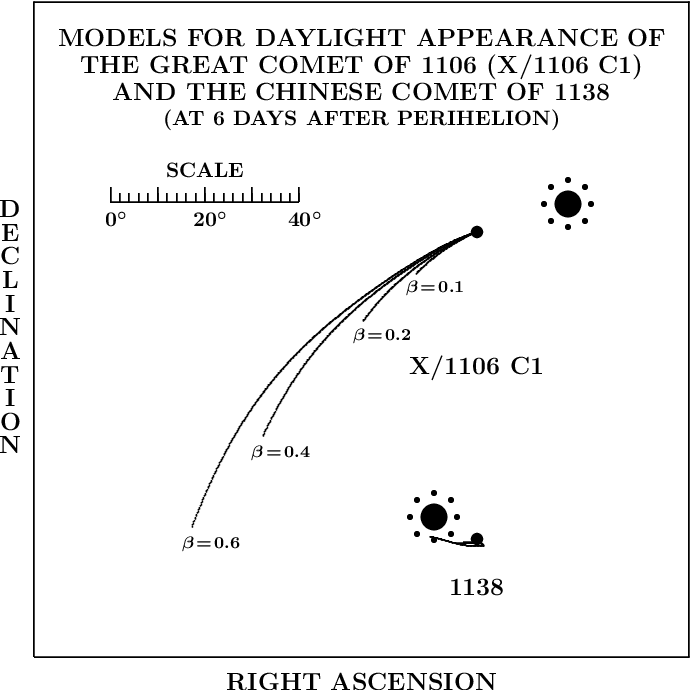}}}
\vspace{0cm}
\caption{Dust-tail development in the 1106 and 1138 comets 6~days
after perihelion (1106 February~8 for the first object, 1138 August~7
for the second one) in projection onto the plane of the sky.  The
tails are represented by the syndynames for radiation pressure
accelerations \mbox{$\beta = 0.1,\,0.2,\,0.4,$ and 0.6} the solar
gravitational acceleration.   Dust particles of all ages up to
5.97~days are plotted.  The contrast between the two objects is
stunning.  Whereas the 1106 comet managed to build up a tail about
100$^\circ$ in length (visible in and beyond twilight) in the
southeasterly direction in the equatorial coordinate system,
the tail of the 1138 comet was extremely short and strongly bent.
It started to the southwest, away from the Sun, but about 1$^\circ$
from the nucleus it turned sharply toward the Sun, terminating a
few degrees to the south of it.  This hook-like shape was entirely
a projection effect, which severely restrained the tail's visibility
for the terrestrial observer.  In reality, the tail rapidly receded
from the Earth at all times.  Note that in projection, the 1106
comet was about twice as far from the Sun as the 1138 comet at a
given time after perihelion (cf.\ Table~2).  The size of the Sun
is not drawn to scale.{\vspace{0.6cm}}}
\end{figure}
Based on the constrained radiation-pressure accelerations of grains,
the dust tails of the 1106 and 1138~comets are modeled schematically in
Figure~4, which~displays the tails six days after perihelion with an identical
system of syndynames projected under the early February~and early August
conditions.~The difference~is~\mbox{astonishing}.  In February, when
the tail debris was expanding through space in a general direction of the
Earth, the 1106 comet managed to build a spectacular tail that extended
over half of the sky.  On the other hand, in August, when the motion of
the 1138 comet was generally away from the Earth, the projected tail
exhibited very sharply bent contours.  It was pointing away from the Sun
near the head, but its orientation then reversed by nearly 180$^\circ$,
in the direction of the Sun, in the proximity of which the feature
terminated.  The hook-like shape was entirely a projection effect that
contributed to the comet's lackluster appearance at the time it was
supposed to be at its brightest.

\begin{table*}[t] 
\vspace{0.17cm}
\hspace{-0.2cm}
\centerline{
\scalebox{1}{
\includegraphics{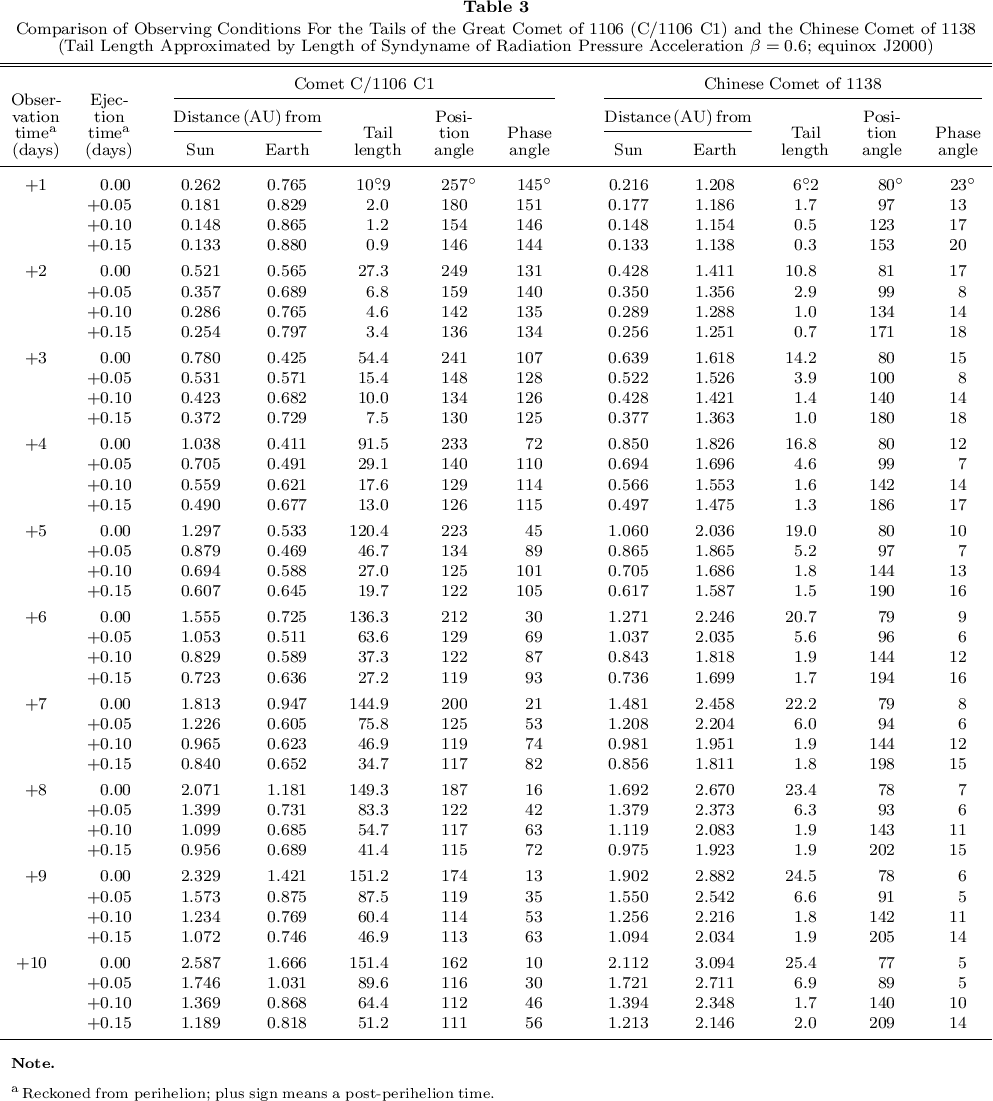}}}
\vspace{0.6cm}
\end{table*}

The enormous disparity between the properties of the projected dust
tails of the 1106 and 1138 comets in the early post-perihelion period
is apparent from Table~3, in which I show, as a function of observation
time, the heliocentric and geocentric distances of particles subjected
to radiation pressure acceleration of 0.6 the solar gravitational
acceleration; their projected distance (as a tail length) and direction
(as a position angle for equinox J2000) from the nucleus; and their
phase angle.  It should be pointed out that the ejecta released within
a few tens of minutes of perihelion are subjected to strong sublimation
effects and do not contribute much to the post-perihelion tail, as
confirmed by the orientations of the observed tails.  Comparison with
the position angle of the radius vector in Table 2 shows, for example,
that one day later the perihelion ejecta of the 1106 comet projected
in a direction only 50$^\circ$ away from the sunward direction.
At different times after perihelion, dust of different ejection times
contributed to the tail differently,~but the tail length was determined
by the ejecta that left the nucleus only a fraction of the day after
perihelion.

\begin{figure}[t] 
\vspace{0.17cm}
\hspace{-0.2cm}
\centerline{
\scalebox{0.59}{
\includegraphics{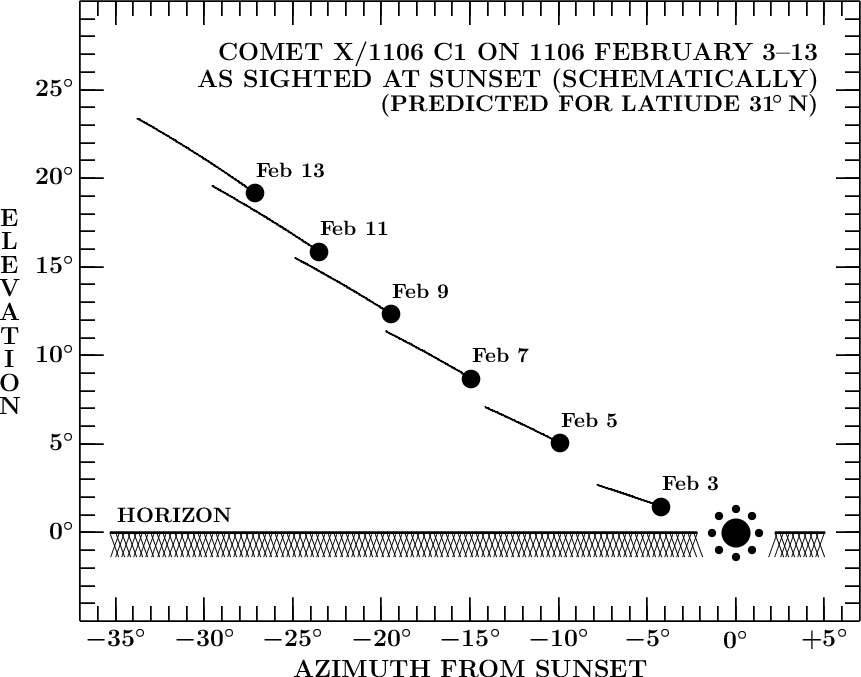}}}
\vspace{0cm}
\caption{The positions of comet X/1106 C1 in the sky at sunset on
 1106 February~3--13 at the geographic latitude 31$^{\:\!\!\circ}$N
 and the longitude 120$^\circ$E, an approximate location of
 the capital of the Sung Dynasty (near today's cities of Hangzhou
 and Nanjing).  The Sun's azimuth at sunset was about 75$^\circ$,
 measured from the south westward.  The plotted tail arcs illustrate
 the feature's orientations, not lengths.  The Sun's dimensions are
 strongly exaggerated.{\vspace{0.5cm}}}
\end{figure}

\section{Comet X/1106 C1 at and After Sunset{\vspace{-0.08cm}}} 
Continuing with the investigation of the observing conditions in
twilight, I plot in Figure~5, in the horizontal coordinate system,
the positions of the Great Comet at sunset between 1106 February~3
and 13.  With the Chinese observations in mind, the plot refers to
a longitude and latitude of, respectively, 120$^\circ$E and
31$^{\:\!\!\circ}$N, on the assumption that the comet was at perihelion
on February~2.25~TT.  Over the ten days, the sunset times varied
from 9:40 to 9:48~UT, the solar elongations of the comet from
4$^\circ\!$.5 to 33$^\circ\!$.3, and its elevations from
1$^{\:\!\!\circ\!}$.5 to 19$^\circ\!$.2.

Although the head of the comet was not bright enough to be detected
at sunset, Figure~5 shows that the observing conditions at the
northern latitudes near 30$^\circ$ were rapidly improving with time
in early February.  This applies not only to the Chinese observations,
but also to those from Palestine (e.g., the city of Jerusalem is at
nearly the same latitude) and other locations.  The large recorded
disparities in the tail length could be explained by the wide variety
in both the sites and times of observation.  Reports of shorter tail
lengths could be expected from more northerly locations and from later,
post-twilight observations, when the comet's head was deep below
the horizon and only distant parts of the tail could still be seen.
This means that the best estimates of the tail length were the
longest ones, near 100$^\circ$.  The reports that the comet was
``split into many pieces,'' or ``appeared like a broken-up star,'' or
like ``a beam forking into many rays,'' provide likely evidence for
the presence of a plasma tail and its complex morphology.~Although~in
the equatorial coordinate system the predicted position of the plasma
tail was in the east-southeast direction (varying from position angle
of 116$^\circ$ on February~5 to 103$^\circ$ on February~13), in the
horizontal system at latitude 31$^{\:\!\!\circ}$N the tail should
have been seen to point up and east, deviating at sunset from the
zenith direction by 63$^\circ$ on February~5 and by 55$^\circ$ on
February~13.

\begin{figure}[t] 
\vspace{0.17cm}
\hspace{-0.2cm}
\centerline{
\scalebox{0.58}{
\includegraphics{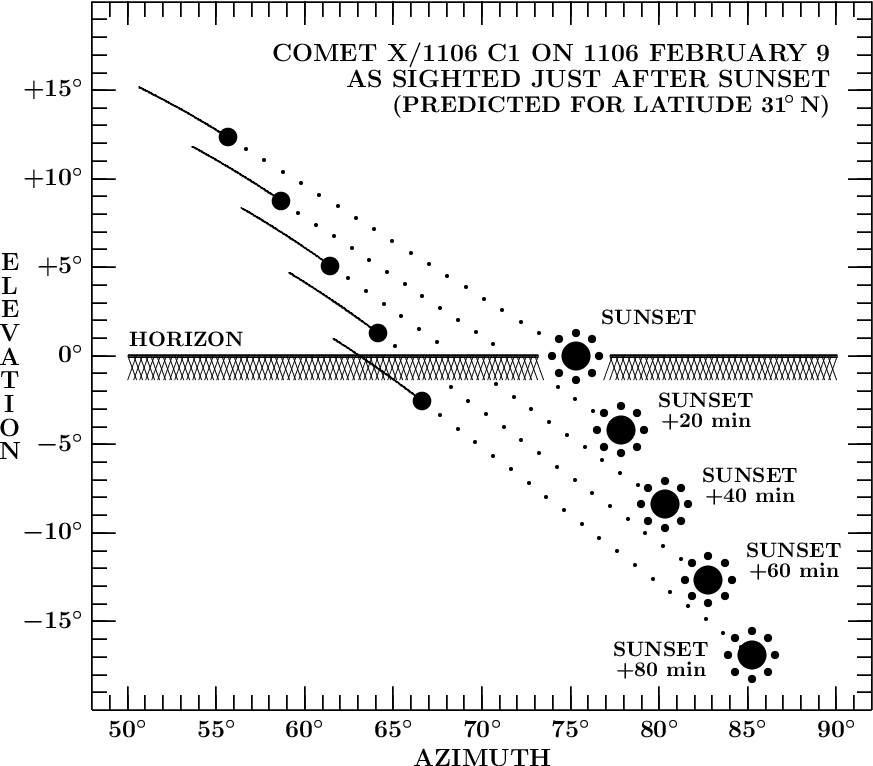}}}
\vspace{0cm}
\caption{Comet X/1106 C1 on 1106 February~9 over a period of
80~minutes, as seen from a location of geographic longitude 120$^\circ$E
and latitude 31$^{\:\!\!\circ}$N.  Sunset is predicted to have occurred
at 9:45~UT, when the comet was 23$^\circ\!$.1 from the Sun at
position angle of 108$^\circ$ and elevation of 12$^\circ\!$.4. The
comet set 67~minutes~later, in the course of astronomical twilight,
with the Sun 14$^\circ$ below the horizon.{\vspace{0.55cm}}}
\end{figure}

To depict the observing conditions~as the 1106 comet was
approaching the horizon following the sunset, I plot in Figure~6 its
apparent motion over a period of 80~minutes on the evening of
February~9, predicted for a location at geographic longitude of
120$^\circ$E and latitude of 31$^{\:\!\!\circ}$N.  At sunset, which
was calculated to have taken place at 9:45~UT, the comet was
23$^\circ\!$.1 from the Sun at position angle of 108$^\circ$ and
elevation of 12$^\circ\!$.4.  The~comet~set 67~minutes later, in
the course of astronomical twilight, with the Sun 14$^\circ$
below the horizon.  Under these~conditions~the long tail must have
offered a magnificent~spectacle.

\begin{table}[b] 
\vspace{0.55cm}
\hspace{-0.2cm}
\centerline{
\scalebox{1}{
\includegraphics{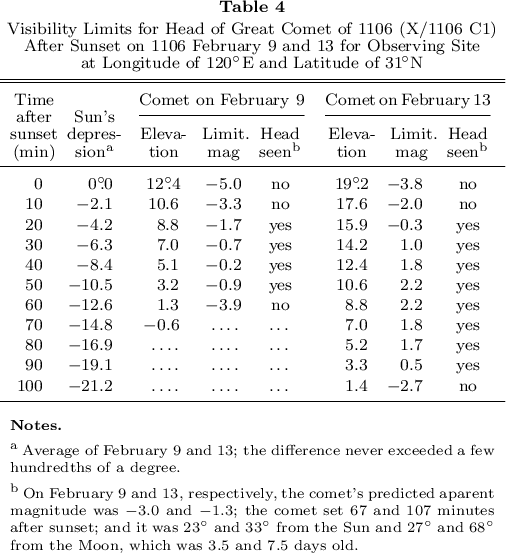}}}
\vspace{-0.08cm}
\end{table}

The potential visibility of the comet's head in twilight was
another issue of interest.  I already noted that the head was too
faint to detect at sunset.  As the Sun was moving deeper below
the horizon, the conditions were improving.  Unfortunately, the
comet's elevation was also diminishing.  One would expect that
optimum conditions may have taken place somewhere midway between
sunset and setting of the comet, a guess that is supported by
quantitative examination in Table~4.  For the perihelion time
of 1106 February~2.25 and the selected observing site in China,
the post-sunset conditions were compared for February~9 (sunset
at 9:45~UT) and February~13 (sunset at 9:48~UT). The predicted
apparent magnitudes of the comet were $-3.0$~mag and $-1.3$~mag,
respectively, and the effects of interference by the Moon were
accounted for; it was 3.5~days old and 27$^\circ$ from the comet's
head on February~9, 7.5~days old and 68$^\circ$ away on the 13th.
The table shows that the comet's head is predicted to have been
visible to the naked eye over a period of 45~minutes (from 12 to
57~minutes after sunset) on February~9 and over a period of
81~minutes (from 14 to 95~minutes after sunset) on February~13.
At the most favorable times on the two days the comet's head is
found to have been, respectively, 2.8~mags and 3.5~mags above
the naked-eye visibility threshold.

In summary, the present examination suggests that the {\bf sungrazing}
orbit for the Great Comet of 1106 offered {\bf favorable} conditions for
its observing both in {\bf daylight} at close proximity to the Sun in
the first three days following the perihelion passage and under {\bf
twilight} conditions on subsequent days, when sightings of the comet's
head were constrained to fairly short intervals of time but the {\bf
long, brilliant tail} --- apparently consisting of both the plasma and
dust constituents --- was what made the {\bf comet great}.  The favorable
geometry contributed significantly:\ the comet approached the Earth to
0.8~AU about one week after perihelion and parts of the dust tail to
less than 0.5~AU a few days earlier.  The small geocentric distances made
the {\bf tail look larger}, as well as the {\bf head and tail brighter},
in the earliest post-perihelion days also because of strong effects of
forward scattering by microscopic dust.  The bulk of the~comet's~recorded
{\bf observations}, made in daylight, twilight, and nighttime, are {\bf
consistent with} the results of this exercise, supporting the suggestion
that this object indeed was a {\bf Kreutz sungrazer}.

My final point on the 1106 comet concerns~the~time~of perihelion, for
which Hasegawa \& Nakano (2001)~found January~26\,$\pm$\,5, following
their reexamination of an earlier result, January~30\,$\pm$\,5, by
Hasegawa\,(1979).~The~large error notwithstanding, Hasegawa \& Nakano's
perihelion time could not fit the Gembloux observation, because the
comet would no longer be in close proximity of the Sun on February~2.
In fact, the time spent by the object, before {\it and\/} after
perihelion, within 1--2$^\circ$~from~the~Sun is seen from Figure~2
to have been shorter than 24~hours, which essentially rules out all
perihelion times before about February~2.0~TT.

\section{The Comet of 1138 at and before Sunrise} 
Turning now to Population\,II,\,the reader will remember my conclusion
from Section~3.3 that the comet~of~1138 was too faint to detect in
daylight shortly after~perihelion.  The plasma tail could not then
be observed and the appearance of the dust tail certainly did not
contribute anything to facilitate detection (Figure~4).

\begin{table}[t] 
\vspace{0.17cm}
\hspace{-0.2cm}
\centerline{
\scalebox{1}{
\includegraphics{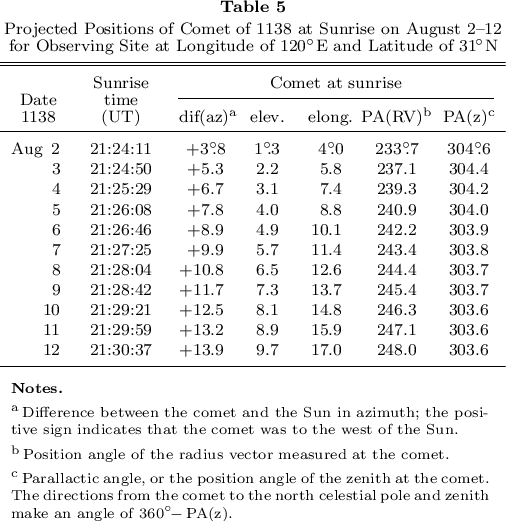}}}
\vspace{0.5cm}
\end{table}

The 1138 comet was located to the west of the~Sun~as it was emerging
from perihelion, becoming eventually~a morning object.  The comet's
very small solar elongations, barely 15$^\circ$ some 10~days after
perihelion, shown in Table~2, did not offer much optimism, but the
issue not addressed by the table was the comet's position relative
to the Sun in the horizontal coordinate system.  Examining the comet's
location at sunrise, I now followed an avenue of inquiry similar to
that employed for the 1106 comet.  The results, again applicable to
the selected Chinese site, turned out to be extremely disappointing
(Table~5).  Fully 11~days after perihelion the comet was less than
10$^\circ$ above the horizon when the Sun already was at the horizon.
Under such conditions, there was no hope whatsoever to pick up the
comet with the naked eye as a twilight object in the early
post-perihelion days.

What happened then that only three weeks later, in early September,
the comet did get discovered?  The answer has three parts:\ (i)~a
gradually increasing solar elongation that prompted the visibility
threshold to drop much more rapidly with time than was the fading
rate~of the comet; (ii)~an intervening event increasing temporarily
the tail's surface brightness; and (iii)~favorable tail-projection
developments.  I now comment on each point.

\begin{table*}[t] 
\vspace{0.17cm}
\hspace{-0.22cm}
\centerline{
\scalebox{1.008}{
\includegraphics{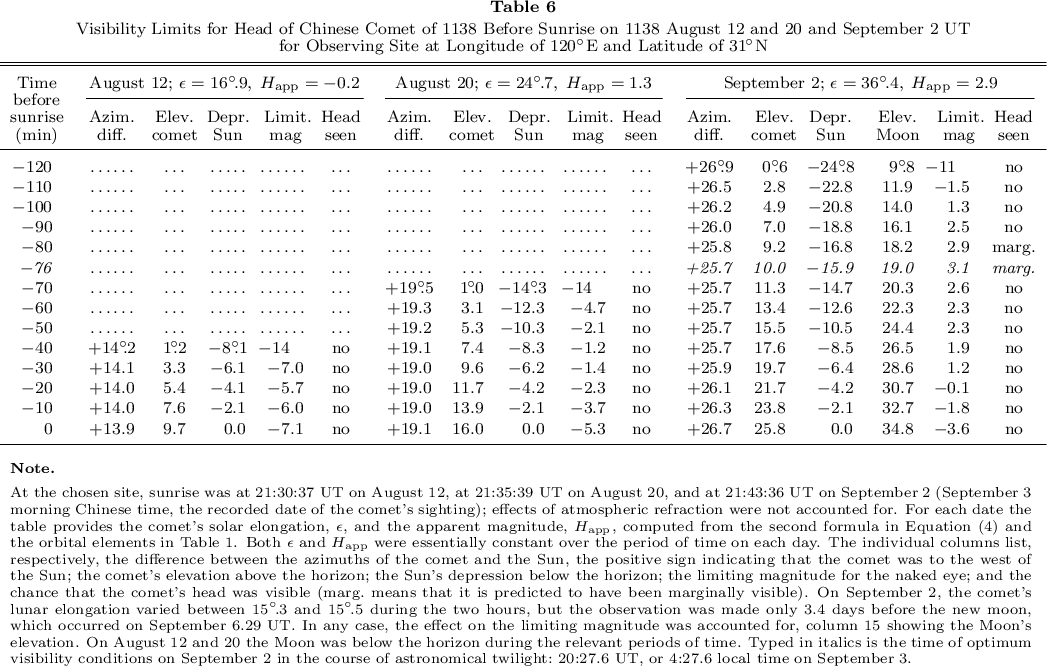}}}
\vspace{0.49cm}
\end{table*}

The best way to document the first point is to compare the observing
conditions at a few relevant times.  I choose the above date of
August~12 and the all-important date of September 2 (UT, i.e.,
September~3 local time) --- the day of observation.  In addition,
August~20 was selected as the third date, because the Earth was then
crossing the comet's orbital plane.  Focusing first on the comet's
head, Table~6 shows the relation between the predicted apparent
magnitude and the limiting magnitude.  On August~12 the comet is
predicted to have been at least 5.5~mags fainter than the visibility
threshold, on August~20 at least 2.5~mags fainter.  In early September,
at the time of recorded observation, the comet's brightness is predicted
to have been comparable to the limiting magnitude\footnote{It is noted
that these numbers do differ a little from those in Sekanina \& Kracht
(2022) because of slightly different laws for the light curve used,
now including the phase term.} over a short period of time, about
midway between the rise of the comet and sunrise.  The comet's head
is predicted to have been at this time slightly brighter than magnitude
3, about 10$^\circ$ above the horizon, and 36$^\circ$ from the Sun,
which was some 16$^\circ$ below the horizon.  The computed limiting
magnitude is of course subject to rather large uncertainties for a
number of reasons, so that the head may of may not have been seen in
early September.  There is in fact a good chance that it was not,
the argument being offered below.

The recorded sighting is that of a {\it hui\/} comet, that is, one
with a tail.  On September~2 the comet already was 1.76~AU from the
Earth, but only 1.12~AU from the Sun, so it is likely that both the
plasma and the dust tails were present.  The syndynames in the dust
tail for September~2.8525~UT (shown in italics in Table~6) are
displayed in Figure~7, which suggests that, compared~to~August~7
(Figure~4), the tail's projected figure remained essentially unchanged,
except for its considerable expansion.

Of distinct assistance to the dust tail's improved visibility was the
Earth's transit across the comet's orbital plane on August~20, increasing
the tail's surface brightness; this was the intervening event under (ii).
As a result, the line of sight from the Earth made only 4$^\circ\!$.7
with the orbital plane at the time of observation on September~3. Also,
because of the orbital plane crossing, the tail was curving in the
opposite sense in early September (Figure~7) compared to August~7
(Figure~4).

\begin{figure}[b] 
\vspace{0cm}
\hspace{-0.2cm}
\centerline{
\scalebox{0.57}{
\includegraphics{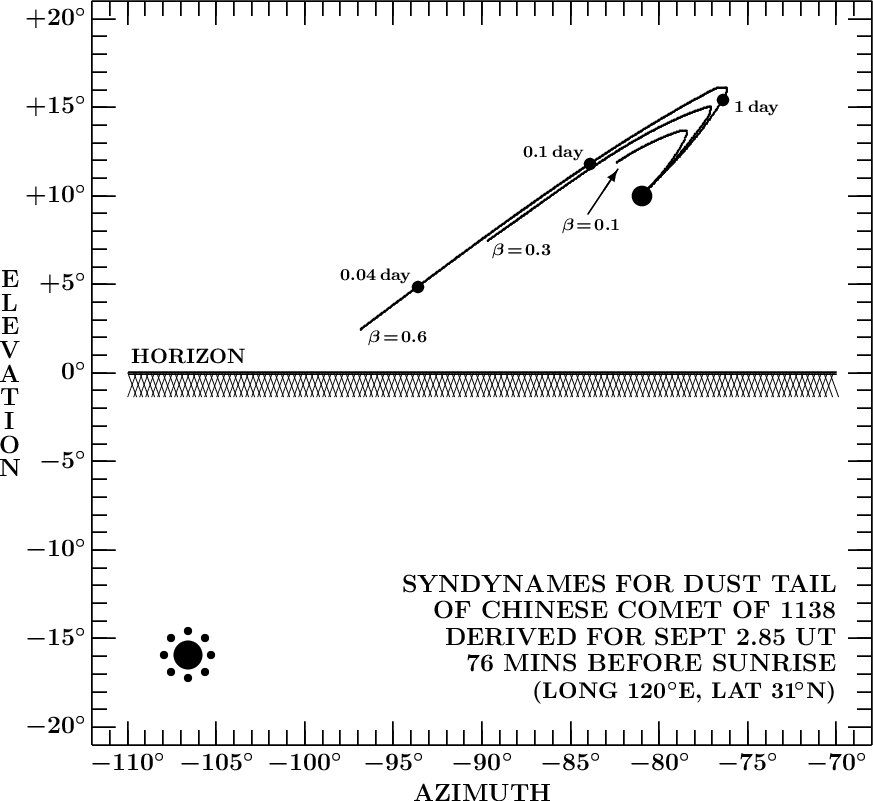}}}
\vspace{-0.03cm}
\caption{Syndynames for \mbox{$\beta = 0.1,\;0.3,\;{\rm and} \; 0.6$} in
the dust tail of the 1138 comet on September~2.85~UT, in the horizontal
coordinate system.  Azimuth is reckoned from the local south to the west,
so that the east is at $-$90$^\circ$.  The comet's head is the large
solid circle at elevation 10$^\circ$, the three small circles on the
syndyname \mbox{$\beta = 0.6$} mark the debris ejected, respectively,
0.04, 0.1, and 1~day after perihelion.  The Sun is shown near 16$^\circ$
below the horizon in the lower left.  The sharp bend in the tail was
persisting, entirely a result of projection.  Ideally, the tail's length
could extend over 25$^\circ$, but a more realistic estimate is about one
half of it.  The comet's head probably was not visible, because the
medieval observer would be puzzled by its unusual position in the tail,
commenting on it.{\vspace{0cm}}}
\end{figure}

Because all dust ejected later than about 0.1 day after perihelion is
seen in Figure~7 to have by early September been projecting at larger
distances from the Sun and at greater elevations than the comet's head,
it was this part of the tail that had the best chance of detection at
that time.  This is my argument under point (iii) above.  In
particular, one can see from the figure that the dust ejected near
1~day after perihelion was in fact located at an elevation about
6$^\circ$ higher in the sky than the head.  It is hard to estimate
how close to the horizon the feature could have been followed, but
chances were that the visible tail length was at most 10$^\circ$ or
so.  It is also likely that the predicted marginal naked-eye detection
of the head (Table~6) meant in practice that it did not stand out enough
to be readily noticed.  Its position, sideways from the main body of
the dust tail and near the middle along its long axis, was very unusual
and to the medieval observer puzzling to the point that, if seen, he
probably would have commented on it.

\begin{table}[b] 
\vspace{0.6cm}
\hspace{-0.2cm}
\centerline{
\scalebox{1}{
\includegraphics{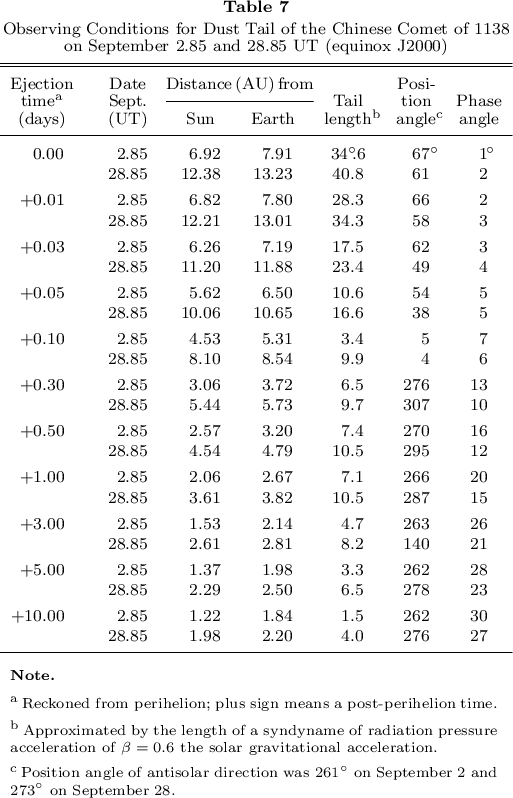}}}
\vspace{0.13cm}
\end{table}

The highly unusual conditions of tail projection are likewise underlined
by the data presented in Table~7, which further assist in estimating
the tail length in early September and also help explain why the comet
was seen over a rather limited period of time.  The striking features
of the table are the enormous distances from the Sun, and even more so
from the Earth, reached by the submicron-sized dust particles (subjected
to a radiation pressure acceleration of 0.6 the solar gravitation
acceleration) ejected in the first hours after perihelion.  (Such
grains dominated the tails of the great Kreutz sungrazers, such as
Ikeya-Seki, early after perihelion.)  For example, particles ejected
a little less than one hour post-perihelion, which in Figure~7 make up
the part of the tail appearing just barely above the horizon, were
located more than 6~AU from the Sun and more than 7~AU from the
Earth on September~2~UT and between 11 and 12~AU on September~28~UT,
i.e., the morning hours of September~29 local time, when the comet
went out of sight (Ho 1962).  On the other hand, the heliocentric
and geocentric distances of the microscopic dust ejected as late as
24~hours after perihelion, which occupies the area of the tail near
its highest point in Figure~7, were only between 2 and 2.7~AU on
September~2 and between 3.6 and~3.9~AU~on September~28; of course,
the production rate of dust was much higher one hour after perihelion
than 23~hours later.

The presence of the plasma tail should have further facilitated
sightings of the comet in early September.  If the plasma tail
extended essentially along the radius vector, the relation between
its angular length $\Theta$ and linear length $\Lambda$ was a
function of the comet's geocentric distance $\Delta$ and the phase
angle $\alpha$:
\begin{equation}
\tan \Theta = \frac{\Lambda \sin \alpha}{\Delta \!+\! \Lambda \cos
 \alpha}.
\end{equation}
With \mbox{$\Delta \!=\! 1.76$ AU} and \mbox{$\alpha \!=\! 32^\circ\!$.1}
on September~2.85~UT, one gets \mbox{$\Theta \!\simeq\! 7^\circ$} for
a tail 0.5~AU long and \mbox{$\Theta \!\simeq\! 11^\circ\!$.5} for a tail
1~AU long.  It therefore cannot be ruled out that the plasma tail appeared
a little longer than the dust tail in the antisolar direction, reaching
perhaps up to nearly 5$^\circ$ farther above the horizon than the dust
tail.  By late September the heliocentric distance increased to almost
1.7~AU and the plasma tail should have been waning and with the
geocentric distance approaching 2~AU, its length should have been
curtailed, making its detection no longer likely.
 
These considerations and results of the modeling of observing
conditions for the comet of 1138 lead to two important conclusions
that revise the currently recognized notions about the Kreutz
sungrazers reaching perihelion at the ``wrong'' time of the year
(Section~2).  It is likely that the conclusions apply to the entire
period from mid-March until mid-August, even though they strictly
refer only to the weeks around August~1.

The {\bf first conclusion} covers the problem of detection.  A Kreutz
sungrazer reaching perihelion at the ``wrong'' time of the year
cannot be discovered with the naked eye in broad daylight near the Sun,
at its brightest, because it is not bright enough.  On the other
hand, there appears to be a narrow window of opportunity to see
such a comet with the naked eye {\it even at low to moderate
northern latitudes\/} starting about one month after perihelion,
as a morning object low above the eastern horizon.  The optimum
conditions are strongly constrained not only in terms of the observing
period (a few weeks),\footnote{The severity of this limitation
depends on the fading rate of the comet, for the northern-hemisphere
observers it is also affected by the comet's modest southerly motion.
The declination of the 1138 comet was close to $-$2$^\circ\!$.5
on September~2, but close to $-$10$^\circ\!$.5 on September~28
(equinox of the date).\vspace{0.1cm}} but especially in terms of
the time of the day, an interval whose length is curbed by a range
of elevations of the comet allowed between its rise on the one hand
and progressing twilight prior to sunrise on the other hand.

In this context, comparison with the sungrazing comet Pereyra (C/1963~R1)
is illuminating.  It passed perihelion on September~23, merely three
weeks later in the year than predicted for the comet of 1138.  Accordingly,
Pereyra's observing conditions were more favorable, but not by much.  Also,
Pereyra was not a member of Population~II, so that the geometry was slightly
different.  Yet, while intrinsically brighter than Ikeya-Seki,\footnote{From
sets of visual brightness observations I determined the post-perihelion
absolute magnitudes of 5.9 for Ikeya-Seki and 4.4 for Pereyra (Sekanina
2002); these compare with an adopted absolute magnitude of 0.5 for the
comet of 1138.} Pereyra was {\it not detected in daylight\/},\footnote{If
the arrival of comet Pereyra were delayed by 16~years (less than 2~percent
of its orbital period), it would have saturated the imager on board the
Solwind satellite.} but was {\it discovered three weeks after
perihelion\/}.~Over a period of several days~\mbox{following} its discovery,
Pereyra\,{\it was\,a\,naked-eye\,object, \mbox{displaying}~a tail\/} several
degrees in length.  Although the conditions were clearly more favorable in
the southern hemisphere (e.g., Bennett et al.\ 1964),{\vspace{-0.03cm}}
the comet and its tail were seen by {\it northern-hemisphere observers\/}
as well.\footnote{The visibility of comet Pereyra with the naked eye is
appar\-ent from the narratives by some observers [e.g., Capen (1964),~who~was
stationed at Wrightwood, California], but also from~an~\mbox{incredible}
story involving a 1977 letter written by G.\,E.\,D.\ Alcock,~in which ---
as aptly recounted in 2003 by B.\,G.\ Marsden in his lecture given~in
Alcock's memory to an RAS-BAA Pro-Am Discussion Meeting:\ {\it
Meteorites, Meteors, \& Comets\/}, documented by Shanklin~(2004) ---
%
%
he wondered, in retrospect, whether a thin pencil-like~beam~of light
that he had seen low in the sky on 1963 September~12, two days before
comet Pereyra was discovered, happened to be the comet's tail.}
Telescopically the comet was under observation over a period of three
months, with the last observation in mid-December made from the northern
hemisphere.  Thus, notwithstanding the various differences (such as
orbital incompatibility, a gap of three weeks in the perihelion time,
an intrinsic-brightness gap of almost 4~magnitudes, etc.), one notices
very obvious similarities between the predicted traits of the 1138 comet
and the observed traits of comet Pereyra.

The {\bf second conclusion} concerns a rule of thumb to investigate a
correlation between the time of last naked-eye observation and the
comet's apparent magnitude.  Sightings of a bright sungrazing comet
in the ancient or medieval times were those of its tail, which was seen
even after the comet's head was no longer visible to the unaided eye.
I tested this relationship on the well-observed sungrazers, such as
the Great March Comet of 1843, the Great September Comet of 1882,
Ikeya-Seki, etc., and suggested that, based on the experience with
these objects, the post-perihelion tail of a sungrazer was typically
seen with the naked eye until the brightness of the head diminished
to apparent magnitude of approximately 7, thus becoming a telescopic
object long before the final tail sighting (Sekanina 2022).

Applied to seven objects between AD~943 and 1695, listed by Hasegawa \&
Nakano (2001) as potential Kreutz sungrazers (all reaching perihelion
under favorable observing conditions in October or November, similar to
those for comet Ikeya-Seki in 1965), this exercise led to a revision
of their absolute brightness downward, the correction averaging more
than 2~magnitudes.  The outcome was their better agreement with the
absolute magnitudes of the recent Kreutz sungrazers.

A very different case on that same list is the comet of 1041, which
was already briefly commented on in Section~2.  Hasegawa \& Nakano
assigned it an absolute magnitude of $-$1, making it 2~mags
intrinsically brighter than was my estimate for the Kreutz system's
progenitor (Sekanina 2022)!  There appear to be two problems with this
1041 object, the more serious one being that two different comets ---
either one recorded in China as well as Korea --- were almost certainly
involved, as suggested by both Ho (1962) and England (2002).  The first
of the two comets was discovered on September~1 (Hasegawa \& Nakano
2001) and ``went out of sight after more than 20~days'' (Ho 1962).  The
mentioned rule-of-thumb method offered for it an absolute magnitude of
4, which was 5~mags fainter than proposed by Hasegawa \& Nakano.  The
second comet appeared in November, was seen for more than 10~days,
and appears to have been intrinsically fainter.  Both may have been
Kreutz sungrazers, perhaps even fragments of an object that split not
long before 1041.

However, the estimated (post-perihelion) absolute magnitude~4 for the
September comet of 1041 happens to be brighter than Pereyra's by only
0.5~mag or less and too faint for the parent, especially because
circumstantial evidence suggests that the 1041 comet fragmented heavily
at perihelion.\footnote{The barycentric orbital period of the 1041
comet upon its arrival at perihelion was 678~years, but Pereyra's
original orbital period slightly exceeded 900~years, which implied
the existence of one or more fragments in an orbital-period range of
\mbox{$\sim$450--750}~years.  This would reduce the comet's rate
of post-perihelion fading, thus bringing the relationship between
the absolute magnitudes of the 1041 comet and Pereyra closer to
expectation.{\vspace{0.04cm}}}
If so, its brightness should have been almost comparable to the brightness
of the 1106 or 1138 comets.  To achieve this, the time of the last
naked-eye observation should have coincided with the comet's apparent
magnitude having equaled not 7, but about 4.5, close to Hasegawa \&
Nakano's value.  And this is the second problem with the September
comet of 1041.

Returning now to the Chinese comet of {\bf 1138}, I find that around
September~28.9~UT, when the comet went out of sight (see Ho 1962),
its {\bf apparent magnitude} computed from the second
formula in Equation~(4) was --- astonishingly --- {\bf 4.5}!!  A
second comet that defies the rule of thumb with apparent magnitude~7
and a second comet that passed perihelion in the ``wrong'' time zone
of the year.  Given that the apparent magnitude~7 applied to the
sungrazers that arrived at perihelion at ``favorable'' times of the
year, could it be that the apparent-magnitude threshold in the rule
of thumb correlates with the seasonal variations in the observing
conditions because the sungrazers' activity and the surface brightness
of their tails respond to the conditions in a different manner?
Unfortunately, the currently available data are too scanty and
inaccurate for an in-depth investigation of this problem at this
time.

\section{Fragmented Nucleus of the Comet of 1138} 
Even though I believe that the nucleus of the Great Comet of 1106 did
break up\footnote{It is possible that besides the Great March Comet of
1843, at least one other sungrazer --- C/1668 E1 --- was also a fragment
of the 1106 comet (Sekanina 2025).{\vspace{0.11cm}}} in the course of
its perihelion passage, only the fragmentation event experienced by the
Chinese comet of 1138 is modeled and examined here in some detail.
There have been two good reasons:\ (i)~it is virtually certain that the
Great September Comet of 1882 and Ikeya-Seki were products of this event
(cf.\ Marsden 1967), possibly with other objects; and (ii)~the nuclei of
the 1882 comet and Ikeya-Seki were themselves subject to fragmentation
at perihelion, the first of the two splitting into five to six components;
a search for similarities has been desirable.

The multiple nuclei of the 1882 comet, as well as the existence
of split comets whose breakup is known to be of tidal nature but
occurred at close proximity of Jupiter --- such as D/1993~F2
(Shoemaker-Levy) or 16P/Brooks (C/1889~N1) --- suggest that (a)~the
formation of a chain of more than two {\it nearly equidistant\/}
fragments, buried in an oval cloud of debris, is by no means exceptional;
and (b)~as a rule, the principal fragment is typically located near
the chain's middle, rather than at the leading end, as is usual for
nontidally split comets.  The conditions (a) and (b) combine to
become helpful for predictive purposes.  In applications, the
first condition serves as an impetus in a search for conceivable
fragments among potential historical sungrazers, whereas --- once
the principal fragment has been identified (or probably identified)
--- the second condition is useful in assessing a timeline of the
fragments in the chain, with the aim to constrain the range of their
orbital periods as much as possible.  Of particular interest is any
hint of evidence that additional fragments could be expected to
arrive in the future, especially near future.

The purpose of the following exercise is to reconstruct the
possible appearance of the nucleus of the 1138 comet that the
medieval Chinese observer would have seen around 4:30 in the
morning of September~3 local time (or September~2.85~UT), nearly
33~days after the predicted perihelion, if\,---\,hypothetically\,---\,he
had~a~\mbox{telescope}~of, say,{\lapeq}$\!\!$1$^{\prime\prime}$\,resolution
available.  He probably would have seen the two condensations around
the nuclei of the comet of 1882 and Ikeya-Seki,\footnote{There is
no correlation between the mass of a fragment and the brightness
of its condensation at a particular time; a more massive fragment
may be enveloped by a fainter condensation than a less massive
fragment because of large, irregular fluctuations in activity.}
plus possibly additional condensations, if the nucleus had split
into more than two sizable fragments.

The modeling of the post-perihelion, split nucleus of the 1138
comet relies heavily on two postulates:\ (i)~the Great
September Comet of 1882 was the largest surviving mass of the
progenitor's Lobe~II (i.e., of Population~II), a fundamental
premise of the contact-binary model for the Kreutz system
(Sekanina 2021); and (ii)~it is not a coincidence that the
difference between the barycentric orbital periods of Ikeya-Seki
and the Great September Comet of 1882, amounting to 83~years,
nearly equals the width of the gaps between the osculating orbital
periods of the neighboring nuclei of the 1882 comet, averaging
about 95~years.

In the context of the conditions (a) and (b) above,~the adherence
to these postulates led to the conclusion that at close proxinity
to perihelion the Sun's tidal force did, in all probability, split
the nucleus of the 1138 comet not into two, but at least four or
perhaps five or six components (Sekanina 2025).  Chronologically,
such a chain of fragments could have started with X/1702~D1 and continued
with the ``sun-comet'' of 1792 (Strom 2002),\footnote{Because of the
very vague nature of the ``sun-comets,'' this obviously is a weak entry;
if it was not a Kreutz member, a fragment expected to arrive at about
this time may have been missed.} followed by the 1882~sungrazer
(the original mass) and by Ikeya-Seki.  In order that the principal
fragment be located near the middle of the chain, in line with
condition (b), one can argue that at least one additional fragment
should exist beyond Ikeya-Seki, and it should arrive in the near
future.  There is a good chance that this could happen --- with
the uncertainty of at least several years --- around 2027, while the
next one (if any) should be a matter of distant future.

The appearance of the nucleus region of a split sungrazer is essentially
determined by the angular separations between the major, kilometer-sized
and larger fragments, which are buried in an encompassing cloud of debris
made up of smaller fragments down to microscopic dust.  To describe the
feature's overall dimensions, one needs to establish the osculating orbits
of the major fragments.  Since a tidal breakup that involves no significant
momentum exchange among the major fragments has practically no effect
on their angular elements, they remain equal to those of the parent at
the time~of~breakup, as does the perihelion time.~For~the~1138~comet
they~are \mbox{listed in Table~1.~The~two~remaining~elements}~vary~from
\mbox{fragment to fragment, but the perihelion distances}~mar\-ginally, by a
fraction of the nucleus size.  The only important differences are in the
fragments' eccentricities, $e$, or alternatively in their semimajor
axes, $a$.

The starting point in the process of modeling the appearance of the 1138
comet's fragmented nucleus is the relationship between its osculating and
future orbital periods.  From Table~1 the relevant osculating period is
754.4~yr, while taking the 1138 comet as the previous appearance of the
1882 sungrazer, the future barycentric orbital period is
\mbox{1882.7\,--\,1138.6 = 744.1 yr}.\footnote{The actual calculation
of the period has to be done in Julian dates (not in calendar days),
which are then converted to years.  This is necessary because the
perihelion times in 1138 and 1882 are given, respectively, in the
Julian and Gregorian calendars.  The difference does not show up,
when a result is rounded off to 0.1~year.{\vspace{0.2cm}}}  In terms of $1/a$,
the effect of the planetary perturbations of the 1882 sungrazer's
motion is{\vspace{-0.1cm}}
\begin{eqnarray}
(1/a)_{\rm fut} \!-\! (1/a)_{\rm osc} &\, = \,& 744.1^{-\frac{2}{3}} -
	754.4^{-\frac{2}{3}} \nonumber \\[0.02cm]
 & = & 0.012178 - 0.012067
	\nonumber \\[0.02cm]
 & = & +0.000111\;{\rm AU}^{-1}.
\end{eqnarray}

It is well known that Kreutz sungrazers have no close encounters
with the planets and that their motions are affected only by the
indirect planetary perturbations.  Because the individual fragments
were close enough to one another over the post-perihelion period of
time of significant indirect perturbations, I assume that their
motions were perturbed equally, so that Equation~(7) applies not
only to the 1882 sungrazer, but to the other fragments~as well.
This of course is not true for their motions at times just before their
next arrival, but by then the perihelion time has essentially been fixed.
The prime driver of the \mbox{$(1/a)_{\rm fut} \!-\! (1/a)_{\rm osc}$}
difference is the heliocentric distance at aphelion, which is
substantially affected by the post-perihelion, not preperihelion,
perturbations.

\begin{figure}[t] 
\vspace{0.17cm}
\hspace{-0.2cm}
\centerline{
\scalebox{0.7}{
\includegraphics{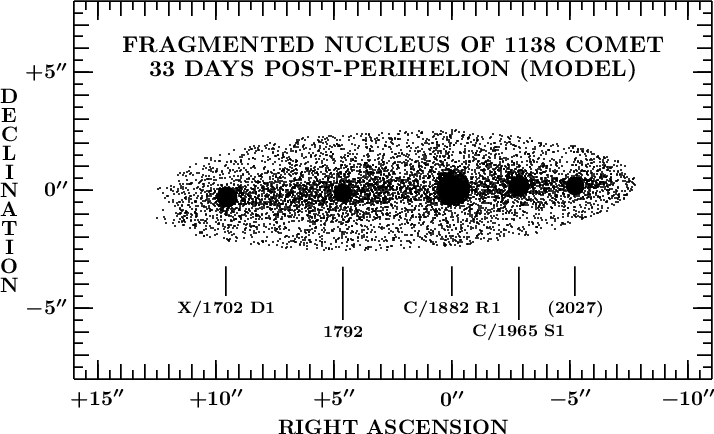}}}
\vspace{0cm}
\caption{Modeled appearance of the fragmented nucleus of the 1138
 comet on 1138 September~2.85~UT, nearly 33~days after perihelion.
 Plotted are the astrometric positions (equinox J2000) of the
 proposed major fragments, besides the Great September Comet of
 1882 and Ikeya-Seki also the possible fragments X/1702~D1 and the
 1792 sun-comet; plus the predicted sungrazer of 2027.  The extent of
 the chain of fragments amounts to 14$^{\prime\prime}\!$.6 in position
 angle of 92$^\circ\!$.3, 77,200~km in space, and is immersed in an
 elongated cloud of dust debris.  All five fragments are assumed to
 be active at the time to a degree that their condensations are not
 overwhelmed by the diffused light from the debris, an ideal case
 discussed below.{\vspace{0.4cm}}}
\end{figure}
Accordingly, $(1/a)_{\rm osc}$ is for each fragment derived from
$(1/a)_{\rm fut}$, which is available with fairly high accuracy from
the consecutive perihelion times, even approximate ones, by applying
Equation~(7).  For example, adopting~for X/1702~D1 a perihelion time of
1702~February~15.1~TT (Kreutz 1901), I get a future{\vspace{-0.04cm}}
orbital period of 563.5~yr, \mbox{$(1/a)_{\rm osc} = 0.014546$
AU$^{-1}\!$}, and \mbox{$e = 0.9998830$}.~\mbox{Similarly}, adopting
a perihelion time of 1792~May~5~TT for the sun-comet of 1792,
1965~October~21.2~TT for Ikeya-Seki, and 2027.7 for the predicted sungrazer
{\vspace{-0.03cm}}(Sekanina 2025), I find \mbox{$(1/a)_{\rm osc} = 0.013165$},
0.011237, and 0.010704~AU$^{-1}$, and \mbox{$e = 0.9998932$}, 0.9999087,
and 0.9999139, respectively.

The last remaining step is to compute the astrometric positions of the
five fragments for the time of interest. Displayed in Figure~8 is
the modeled appearance of the nucleus region of the 1138 comet on
September~3 around 4:30 local time (September~2.85~UT).  The extent
of the fragmented nucleus (from X/1702~D1 to the predicted comet of
2027) amounts to 14$^{\prime\prime}\!$.6 in position angle of 92$^\circ\!$.3.
Terminating the chain of fragments instead with Ikeya-Seki, the length
would be 12$^{\prime\prime}\!$.2, with Ikeya-Seki and the 1882 sungrazer
only 2$^{\prime\prime}\!$.8 apart.  At the geocentric distance of
1.76~AU this was equivalent to merely 3600~km, but because the
line~of~sight made with the line connecting the two objects an
angle of merely 14$^\circ$, they were in space 14,500~km apart at the
time.

\begin{table}[t] 
\vspace{0.15cm}
\hspace{-0.21cm}
\centerline{
\scalebox{0.99}{
\includegraphics{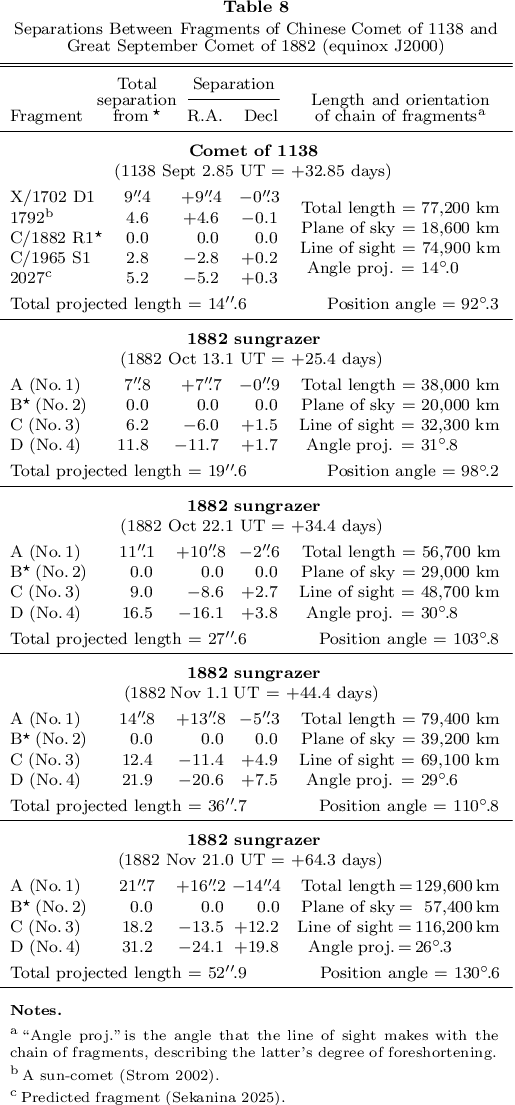}}}
\vspace{0.2cm}
\end{table}

It is illuminating to compare the {\it modeled\/} appearance of the
1138 comet in Figure 8, at a heliocentric distance of 1.12~AU, with
the {\it observed\/}~appearance of the fragmented nucleus of the 1882
sungrazer at similar post-perihelion distances.  Since it will never
be known how the nucleus of the 1138 comet actually looked like, this
exercise is purely academic.  Yet, inspection of the range of possible
appearances offers an insight into the potential subsequent evolution
and ultimate fate of the fragments.

Table~8 compares the two comets in terms of separation distances among
the individual nuclear fragments.  For the 1882 sungrazer, Kreutz (1891)
was able to determine separate sets of orbital elements for four of its
nuclear components by linking the normal places of the preperihelion
nucleus with those of the respective post-perihelion fragment.  The
four condensations, designated A, B, C, and D in the standard
nomenclature, were in the notation used by Kreutz referred to as
points No.~1, No.~2, No.~3, and No.~4, respectively.\footnote{Marsden
\& Williams' (2008) catalogue lists Kreutz's (1891) sets of
nonrelativistic orbital elements for nuclei A, C, and D of the
1882 sungrazer, but Hufnagel's (1919) set of relativistic elements
for nucleus B.  For the sake of uniformity, I use the sets of
nonrelativistic elements by Kreutz for all four nuclei.}

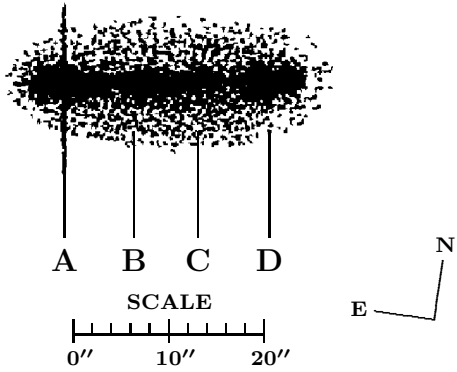
\begin{figure}[t] 
\vspace{0.16cm}
\hspace{-0.4cm}
\centerline{
\scalebox{0.33}{ 
\includegraphics{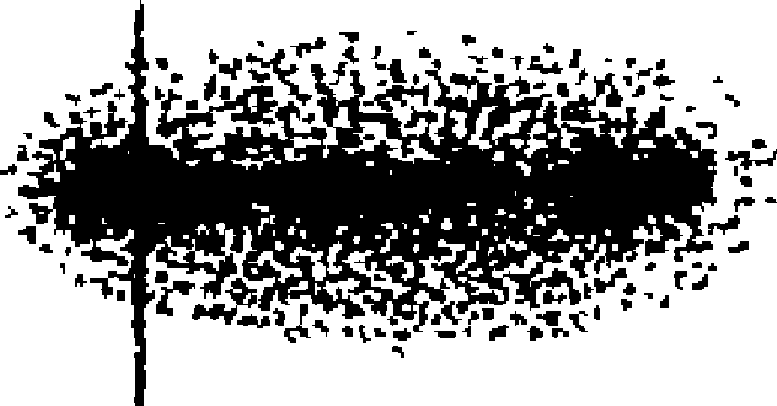}}}

\vspace{-3cm}
\setlength{\unitlength}{1mm}
\begin{picture}(80,50)
\put(26.85,12){\line(0,1){10}}
\put(26.85,9){\makebox(0,0){\large \bf A}}
\put(36.09,12){\line(0,1){14}}
\put(36.09,9){\makebox(0,0){\large \bf B}}
\put(44.61,12){\line(0,1){14}}
\put(44.61,9){\makebox(0,0){\large \bf C}}
\put(54.09,12){\line(0,1){14}}
\put(54.09,9){\makebox(0,0){\large \bf D}}
\put(40.8,3.4){\makebox(0,0){\bf SCALE}}
\put(28,-1){\line(1,0){25.34}}
\multiput(28,-2)(12.67,0){3}{\line(0,1){3}}
\multiput(30.534,-1)(2.534,0){9}{\line(0,1){1.5}}
\put(28,-4){\makebox(0,0){\bf 0\rlap{\boldmath $^{\prime\prime}$}}}
\put(40.67,-4){\makebox(0,0){\bf 10\rlap{\boldmath $^{\prime\prime}$}}}
\put(53.34,-4){\makebox(0,0){\bf 20\rlap{\boldmath $^{\prime\prime}$}}}
\multiput(76,1)(0.0144,0.1){80}{\makebox(0,0){\tiny $\cdot$}}
\multiput(76,1)(-0.1,0.0144){80}{\makebox(0,0){\tiny $\cdot$}}
\put(77.5,11){\makebox(0,0){\bf N}}
\put(66,2.5){\makebox(0,0){\bf E}}
\end{picture}
\vspace{0.6cm}
\caption{Drawing of the fragmented nucleus of the 1882 sungrazer, made
by Elkin on 1882 October 13.1~UT, as seen with a 15-cm Grubb equatoreal.
This was the only time that he was able to resolve four condensations
in the chain; shown are the separations that he reported (Gill et
al.\ 1911).  Kreutz (1888) assigned these positions to nuclei B
through E, but the easternmost fragment was near the very end of the
extended cloud of debris, and it had to be A.  Note that most
condensations are elongated.~The~line crossing the chain was marked
by Elkin to show the point that~was measured for position.  The
principal fragment was B, whose orbital period essentially equaled
that of the nucleus before fragmentation.{\vspace{0.6cm}}}
\end{figure}

The four data sets for the 1882 sungrazer in Table~8 refer to a
selected set of the astrometric and/or physical observations by
W.~L.~Elkin at the Cape Observatory (Gill et al.\ 1911), which
were accompanied by dot renditions of the nuclear chain immersed
in a cloud of debris.  They were drawn to scale.

The first of these reproductions, in Figure~9, shows the fragmented
nucleus on 1882~October~13~UT, 25.4~days after perihelion, as seen
by Elgin with a \mbox{15-cm} Grubb equatoreal.  The extent of about
20$^{\prime\prime}$ exceeds by 5$^{\prime\prime}$ the length of the
modeled chain of fragments of the 1138 comet.  As Table~8 explains,
this apparent imbalance happens to be the effect partly of a larger
geocentric distance of the 1138 object, partly of its highly
unfavorable projection, as the chain of fragments made an angle of
only 14$^\circ$ with the line of sight.  In reality, the 1882
fragmented nucleus was 38,000~km long at the time, less than one
half the length of the modeled 1138 nucleus.

\begin{figure}[t] 
\vspace{0.2cm}
\hspace{-0.17cm}
\centerline{
\scalebox{0.29}{
\includegraphics{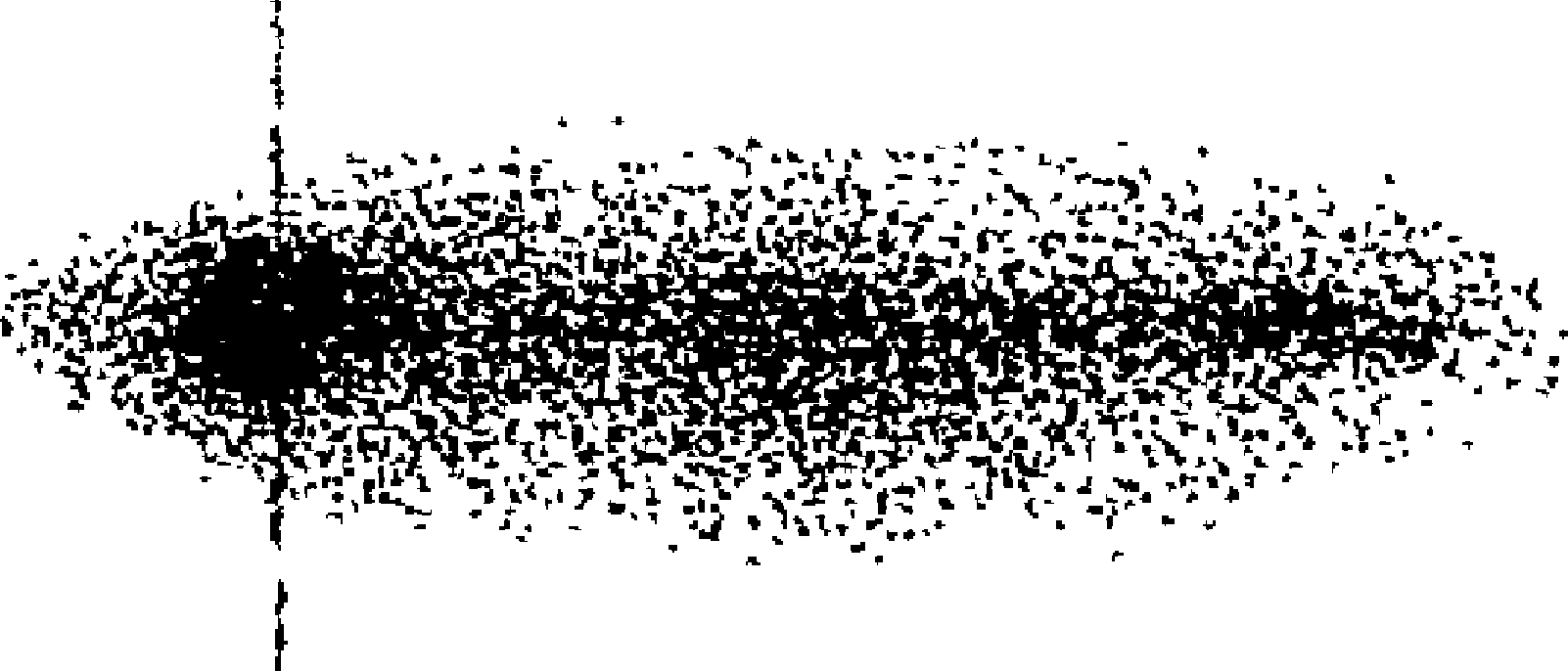}}}

\vspace{-3cm}
\setlength{\unitlength}{1mm}
\begin{picture}(80,50)
\put(17.6,12){\line(0,1){10}}
\put(17.6,9){\makebox(0,0){\large \bf A}}
\put(30.64,12){\line(0,1){14}}
\put(30.64,9){\makebox(0,0){\large \bf B}}
\put(41.22,12){\line(0,1){14}}
\put(41.22,9){\makebox(0,0){\large \bf C}}
\put(50.03,12){\line(0,1){14}}
\put(50.03,9){\makebox(0,0){\large \bf D}}
\put(36.4,3.4){\makebox(0,0){\bf SCALE}}
\put(24.64,-1){\line(1,0){23.5}}
\multiput(24.64,-2)(11.75,0){3}{\line(0,1){3}}
\multiput(26.99,-1)(2.35,0){9}{\line(0,1){1.5}}
\put(24.64,-4){\makebox(0,0){\bf 0\rlap{\boldmath $^{\prime\prime}$}}}
\put(36.39,-4){\makebox(0,0){\bf 10\rlap{\boldmath $^{\prime\prime}$}}}
\put(48.14,-4){\makebox(0,0){\bf 20\rlap{\boldmath $^{\prime\prime}$}}}
\multiput(76,1)(0.02456,0.1){80}{\makebox(0,0){\tiny $\cdot$}}
\multiput(76,1)(-0.1,0.02456){80}{\makebox(0,0){\tiny $\cdot$}}
\put(78.3,10.7){\makebox(0,0){\bf N}}
\put(66,3.5){\makebox(0,0){\bf E}}
\end{picture}
\vspace{0.6cm}
\caption{Drawing of the fragmented nucleus of the 1882 sungrazer, made
by Elkin on 1882 October 22.1~UT, as seen with a heliometer.  Compared
to October~13, condensation A brightened significantly.  The positions
of the fragments B, C, and D are now those calculated from Kreutz's
orbital elements and tied to A.  The overall length of the cloud of
debris is more than twice as long as the distance that separates the
condensations A and D.  Additional, smaller, and gradually disintegrating
fragments (E, F, \ldots) must have been located beyond D.  The line
crossing the chain was marked by Elkin to show the point measured for
position.{\vspace{0.45cm}}}
\end{figure}
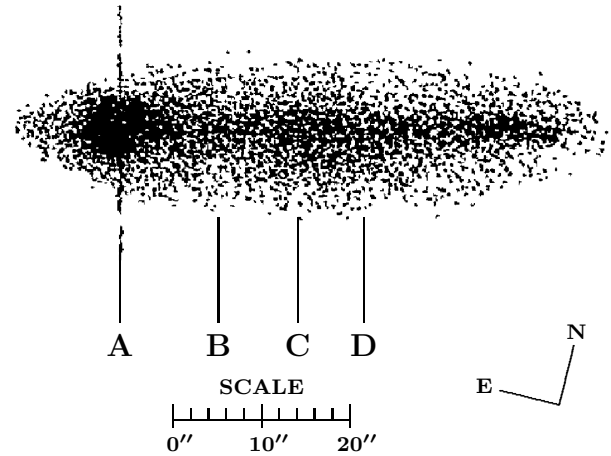

The observation of October 13 was the only one, when Elkin was able to
resolve the nuclear chain into four fairly distinct condensations, so that
the separations among all of them could be measured.  Note however that
most condensations appear to be elongated.  From the east to the west,
the measured distances were 7$^{\prime\prime}\!$.3 between the first and
second condensations, 6$^{\prime\prime}\!$.7 between the second and third,
and 7$^{\prime\prime}\!$.5 between the third and fourth.  Kreutz (1888),
who only saw the report (the drawing was published in 1911), assigned
the condensations to nuclei B, C, D, and E (that is, Nos.~2, 3, 4, and
5 in his notation).  However, Figure~9 clearly shows that the easternmost
condensation was near the very end of the cloud of debris and must have
been nucleus A.

This conclusion is supported by Kreutz's own orbital computations,
which for the three separation distances give, respectively,
7$^{\prime\prime}\!$.8, 6$^{\prime\prime}\!$.2, and 5$^{\prime\prime}\!$.6
(Table~8).  The distances BA and BC agree with Elkin's measurements to
$\pm$0$^{\prime\prime}$.5 and only CD is off by nearly 2$^{\prime\prime}$,
possibly because of a contribution to D from E.  Another strong argument
comes from Elkin's rather extensive description of the easternmost\,(or
following\footnote{In the era of visual observation it was common that
an object to the west of a point of reference be called a {\it preceding
object\/}, one to the east of it a {\it following object\/}, apparently
because when the telescope clock was stopped, the Earth's rotation moved
the star field westward.}) condensation.~Elkin~conveys that ``{\it till
the end of October} [{\it this fragment\/}] {\it was a fair object for
observation.\/''}  However, by November~11 it became ``{\it so covered
with haze, that it was obvious that soon nothing but the centre of the
great mass would be available for a reference point \ldots, and after
Nov.~18 \ldots\,the place of\/} [{\it the following condensation\/}] {\it
was more guessed at than actually seen.\/}''  This is a consummate
account of a {\it protracted flare-up\/} of a fragment that was
undergoing a process of relentless loss of mass, with survival
problems on the horizon.  Certainly not the prospects one would
expect for the principal nucleus B!

To document Elkin's account, I reproduce additional three of his many
drawings in Figures~10--12, which cover the period of October~22
through November~21~UT.  They are annotated to indicate the relative
positions of the nuclei from Table~8, as calculated from Kreutz's
(1891) orbital elements and tied to condensation A (or to what
remained of it on November~21).  Unfortunately, Elkin made these
observations with a heliometer, whose aperture was much smaller than
that of the equatoreal; accordingly, their quality is inferior.

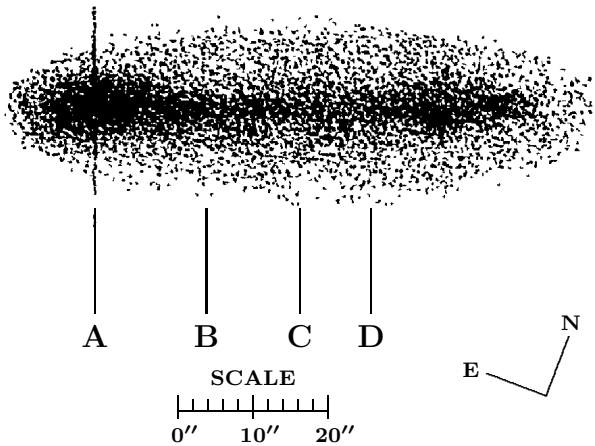
\begin{figure}[t] 
\vspace{0.35cm}
\hspace{-0.171cm}
\centerline{
\scalebox{0.254}{
\includegraphics{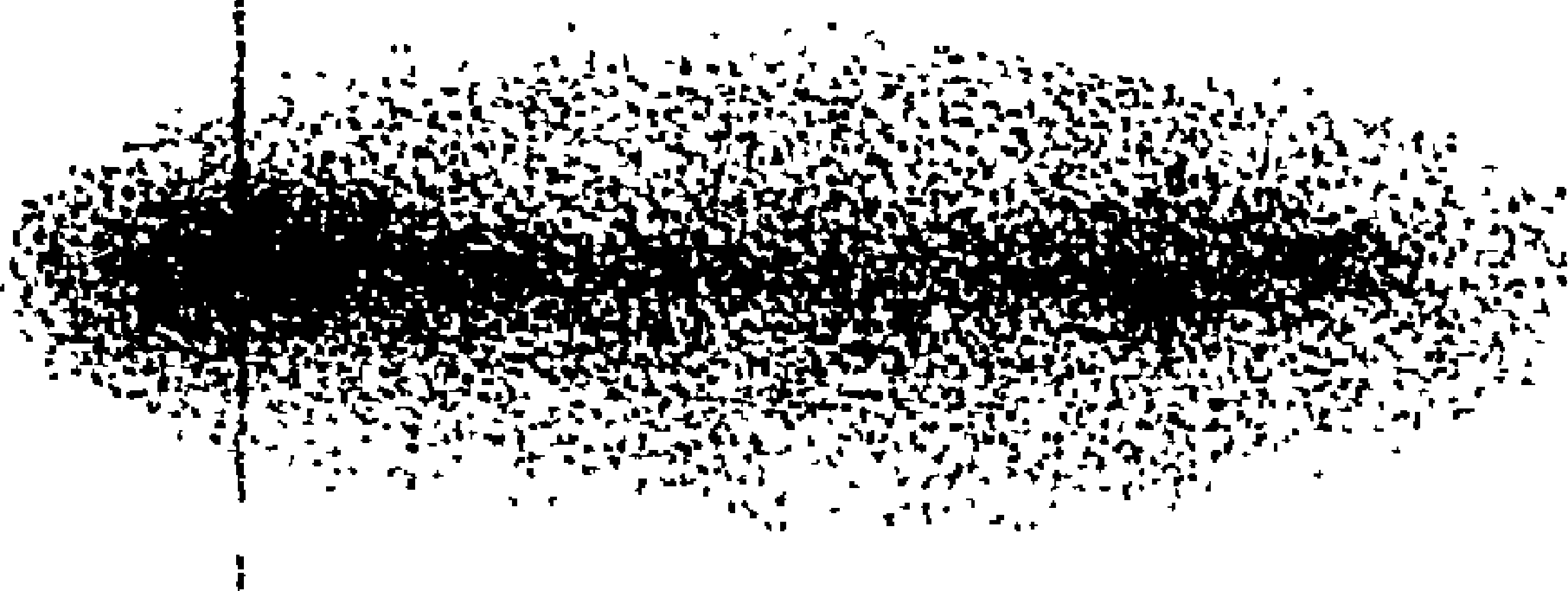}}}

\vspace{-2.7cm}
\setlength{\unitlength}{1mm}
\begin{picture}(80,50)
\put(16,12){\line(0,1){14}}
\put(16,9){\makebox(0,0){\large \bf A}}
\put(30.8,12){\line(0,1){14}}
\put(30.8,9){\makebox(0,0){\large \bf B}}
\put(43.2,12){\line(0,1){14}}
\put(43.2,9){\makebox(0,0){\large \bf C}}
\put(52.7,12){\line(0,1){14}}
\put(52.7,9){\makebox(0,0){\large \bf D}}
\put(37.05,3.4){\makebox(0,0){\bf SCALE}}
\put(27,-1){\line(1,0){20}}
\multiput(27,-2)(10,0){3}{\line(0,1){3}}
\multiput(29,-1)(2,0){9}{\line(0,1){1.5}}
\put(27,-4){\makebox(0,0){\bf 0\rlap{\boldmath $^{\prime\prime}$}}}
\put(37,-4){\makebox(0,0){\bf 10\rlap{\boldmath $^{\prime\prime}$}}}
\put(47,-4){\makebox(0,0){\bf 20\rlap{\boldmath $^{\prime\prime}$}}}
\multiput(76,1)(0.038,0.1){80}{\makebox(0,0){\tiny $\cdot$}}
\multiput(76,1)(-0.1,0.038){80}{\makebox(0,0){\tiny $\cdot$}}
\put(79.3,10.8){\makebox(0,0){\bf N}}
\put(66,4.5){\makebox(0,0){\bf E}}
\end{picture}
\vspace{0.6cm}
\caption{Drawing of the fragmented nucleus of the 1882 sungrazer, made
by Elkin on 1882 November 1.1~UT, as seen with a heliometer.  Compared
to the sketch from October~22 (Figure~10), the changes were relatively
subtle, except for an obvious elongation of condensation A.  While C
temporarily disappeared, B may have now shown up.  Its computed
position would be in good agreement with observation, if the
center of mass of condensation A was moved a little closer to its
eastern boundary than shown by Elkin.  Also apparent is another
condensation about 10$^{\prime\prime}$ to the west of the predicted
position of D, close to where one would expect condensation E (for
which no orbit is available).  The length of the cloud of debris
increased by 10$^{\prime\prime}$ in ten days.{\vspace{0.6cm}}}
\end{figure}

Nonetheless, Figure~10, displaying Elkin's drawing from October~22,
does show major changes that occurred in the appearance of the nucleus
over the nine days.  The most obvious is the dramatic brightening and
expansion of condensation A, suggesting a high rate of its mass loss,
as remarked above.  Another major difference is the considerable
increase in the length of the feature, which, including the cloud of
debris, amounted to 68$^{\prime\prime}$, nearly 2$\frac{1}{2}$ times
the distance between A and D,{\vspace{-0.03cm}} calculated from Kreutz's
(1891) orbital elements.  A slight brightening is seen at the calculated
position of condensation C, but there are no signs of B or D.
As is the case with so many other split comets, no general
relationship exists between fragment size and activity.

Ten days later, the appearance of the comet's chain of fragments did
not look radically different (Figure~11).  The only easily perceived
change was a significant elongation of condensation A, which could
be explained by the presence of microscopic dust in its copious
ejecta, subjected to high solar radiation pressure accelerations.
If so, the center of mass of this condensation should be near its
eastern boundary, and the fainter condensation several seconds of arc
to the west of it would be in very good agreement with the position
of the principal fragment B, derived from Kreutz's orbital elements.
And while there are no signs of fragments C or D, at least one
additional condensation is apparent near the western end of the
chain, at a location where one would expect fragment E.  In the
ten days, the cloud of enshrouding debris increased its extent
from 68$^{\prime\prime}$ to 78$^{\prime\prime}$ and was still
more than twice as long as the computed distance between the
fragments A and D.

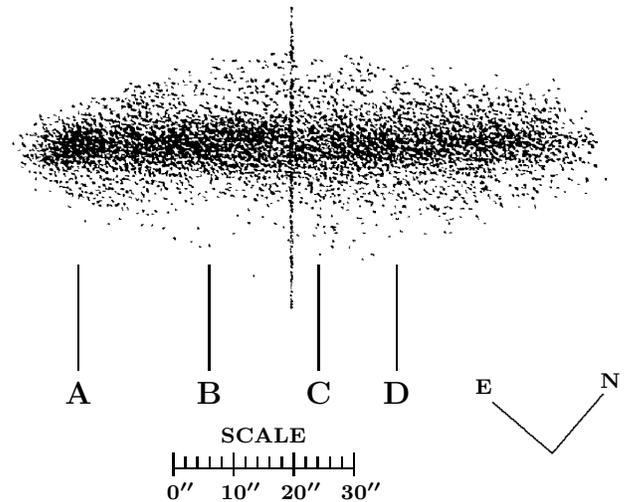
\begin{figure}[t] 
\vspace{0.16cm}
\hspace{-0.1cm}
\centerline{
\scalebox{0.2}{
\includegraphics{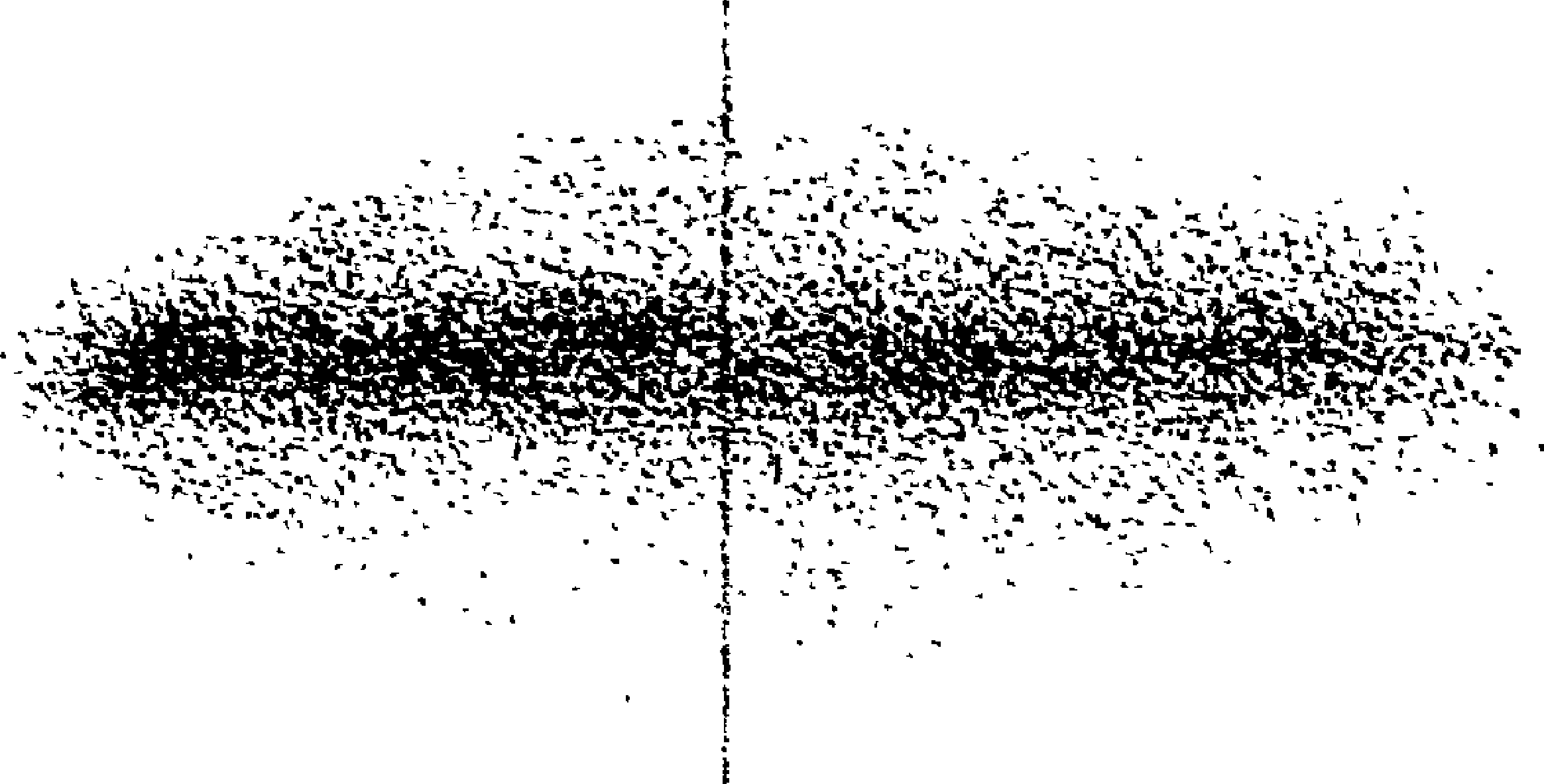}}}

\vspace{-3cm}
\setlength{\unitlength}{1mm}
\begin{picture}(80,50)
\put(13,12){\line(0,1){14}}
\put(13,9){\makebox(0,0){\large \bf A}}
\put(30.36,12){\line(0,1){14}}
\put(30.36,9){\makebox(0,0){\large \bf B}}
\put(44.92,12){\line(0,1){14}}
\put(44.92,9){\makebox(0,0){\large \bf C}}
\put(55.32,12){\line(0,1){14}}
\put(55.32,9){\makebox(0,0){\large \bf D}}
\put(37.6,3.4){\makebox(0,0){\bf SCALE}}
\put(25.6,-1){\line(1,0){24}}
\multiput(25.6,-2)(8,0){4}{\line(0,1){3}}
\multiput(27.2,-1)(1.6,0){14}{\line(0,1){1.5}}
\put(25.6,-4){\makebox(0,0){\bf 0\rlap{\boldmath $^{\prime\prime}$}}}
\put(33.6,-4){\makebox(0,0){\bf 10\rlap{\boldmath $^{\prime\prime}$}}}
\put(41.6,-4){\makebox(0,0){\bf 20\rlap{\boldmath $^{\prime\prime}$}}}
\put(49.6,-4){\makebox(0,0){\bf 30\rlap{\boldmath $^{\prime\prime}$}}}
\multiput(76,1)(0.0857,0.1){80}{\makebox(0,0){\tiny $\cdot$}}
\multiput(76,1)(-0.1,0.0857){80}{\makebox(0,0){\tiny $\cdot$}}
\put(83.7,10.7){\makebox(0,0){\bf N}}
\put(66.9,9.9){\makebox(0,0){\bf E}}
\end{picture}
\vspace{0.6cm}
\caption{Drawing of the fragmented nucleus of the 1882 sungrazer,
made by Elkin on 1882 November 21.0~UT, as seen with a heliometer.
Major morphological changes occurred in the three weeks between
November~1 and 21.  Condensation A faded to the extent that Elkin
was compelled to abandon using it as the point of reference for
positional measurement and employ the middle of the fragmented
nucleus instead.  This drawing illustrates what Elkin meant by
remarking that after November~18 the position of~conden\-sation A
was ``more guessed at than actually seen.'' There are no obvious
features at the computed places of fragments B, C, or~D.{\vspace{0.5cm}}}
\end{figure}

The final of Elkin's four selected drawings of the fragmented nucleus
was made on November~21~UT and is presented in Figure~12.  Major
differences in the feature's morphology over the three November weeks
are readily apparent.  Especially striking is the fading of condensation A,
which prompted Elkin to complain about guessing its place and moving
the point of reference for his astrometric measurement to the middle
of the fragmented nucleus.  The extent of the cloud of debris ballooned
to close to 100$^{\prime\prime}$, while the computed length of the
chain of fragments A--D amounted to some 53$^{\prime\prime}$.  No
measurable features are seen at the computed locations of the fragments
B, C, or D.

This cursory account of Elkin's selected drawings of the fragmented
nucleus of the 1882 sungrazer illustrates the complex evolution of,
and temporal variations in, its appearance.  In the context of the
proposed model for the fragmented nucleus of the Chinese comet of
1138, whose principal component the 1882 sungrazer is believed to have
been, unpredictable chaotic activity is not only inevitable, but in one
sense downright momentous for the individual fragments' future.

I refer of course to a chance of each fragment's survival beyond its
next perihelion as a sizable, essentially intact body.  Regardless of
the model used for the Kreutz sungrazer system, the 1138 and 1882
fragments cannot be readily compared because of the difference of
one generation between them.  In practice, this means that survival
throughout the entire revolution about the Sun a priori is more
likely for the former.  Still, the failure to survive is not ruled
out for either of them.  This uncertainty is a major {\it complication
for predictions\/} of future appearances of Kreutz sungrazers.

What helps is that we know that, in the least, two~fragments of the
1138 comet {\it did survive\/} their following perihelion passage.
And if X/1702~D1 was its fragment as well, as I assume, a total of
at least three fragments survived, because the 1702 sungrazer was
observed after perihelion.  Further surviving fragments may arrive
in the future (Sekanina 2025).

On the other hand, it is anybody's guess, how many (and which)
fragments of the 1882 sungrazer and Ikeya-Seki are going to
survive their next perihelion.  The principal fragments of both
comets and the former's fragment~C (which with B was observed
long into 1883) have the best chance, while fragment A of the
1882 sungrazer, with its highly variable activity, is an excellent
candidate for early disintegration.

\begin{figure*}[t] 
\vspace{0.16cm}
\hspace{-0.2cm}
\centerline{
\scalebox{0.89}{
\includegraphics{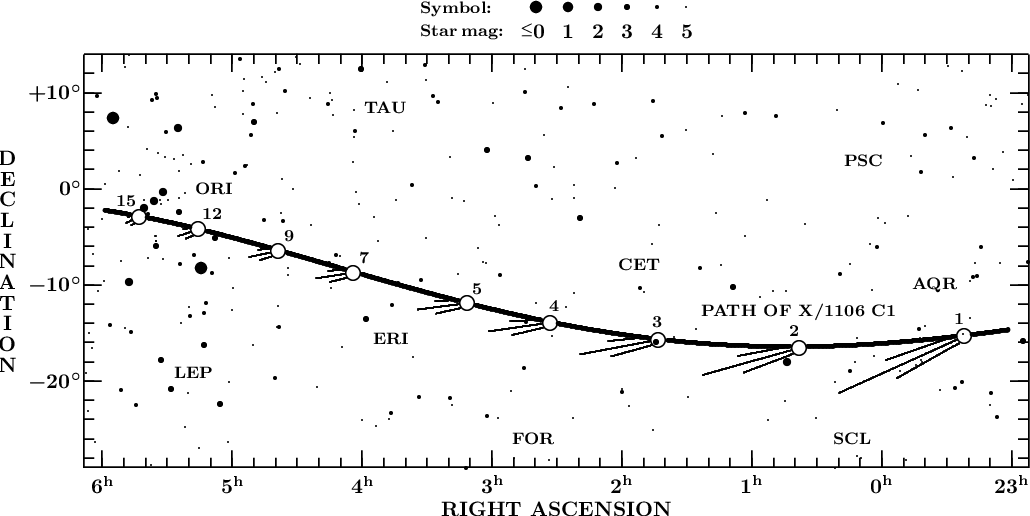}}}
\vspace{0cm}
\caption{Path of the Great Comet of 1106 across the sky (equinox
 J2000).  The positions 1--15 cover the period of time from
 February~6 through April~17 at a 5-day step, each identified
 in Table~9.  The schematically drawn tail only serves to show
 the antisolar direction.  The path, describing the comet's sky
 journey from Aquarius to the belt of Orion, was in its entirety
 seen from much of the northern hemisphere.{\vspace{0.5cm}}}
\end{figure*}

\section{Path Across the Sky and Ephemeris} 
As the representatives of the two foremost populations, the Great
Comet of 1106 and the Chinese comet of 1138 (Ho 403) were on a par
in terms of their significance in the hierarchy of the Kreutz
sungrazer system.  Yet, because of the difference in their timing
on a scale of less than one thousandth of their orbital period,
to the terrestrial observer the first presented a spectacle, the
other a run-of-the-mill ``broom star.''

\begin{figure}[b] 
\vspace{0.5cm}
\hspace{-0.2cm}
\centerline{
\scalebox{0.9}{
\includegraphics{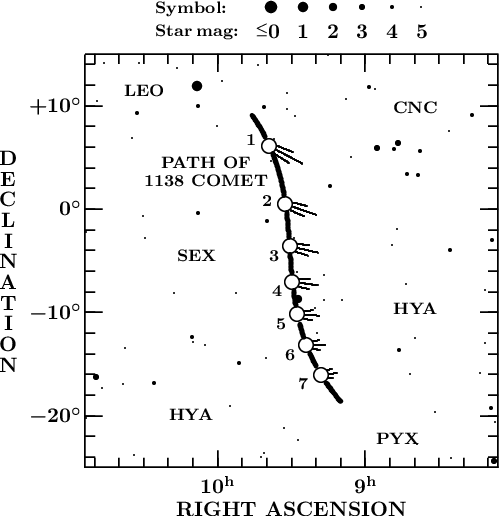}}}
\vspace{0cm}
\caption{Path of the Chinese comet of 1138 (Ho 403) across the sky
 (equinox J2000).  The positions 1--7 cover the period of time from
 August~7 through October~6 at a 10-day step (Table~9).  The tail
 is drawn schematically as in Figure~13.  The comet's path begins
 near Regulus ($\alpha$~Leo) and continues southwards through the
 constellation Hydra, along the boundary with Sextans.{\vspace{-0.06cm}}}
\end{figure}

What were then the paths of the two sungrazers in the sky?  Based on
the orbits in Table~1, this question is answered in Figures~13
and 14.  The ephemerides of the two comets, with the predicted
apparent magnitudes derived from the light-curve laws in (4), are
in Table~9.  The Earth was crossing the comets' orbital planes on
1106 March 7.9 and 1138 August~20.1 TT, respectively.\,\,\,\,\,

\begin{table}[b] 
\vspace{0.5cm}
\hspace{-0.2cm}
\centerline{
\scalebox{0.95}{
\includegraphics{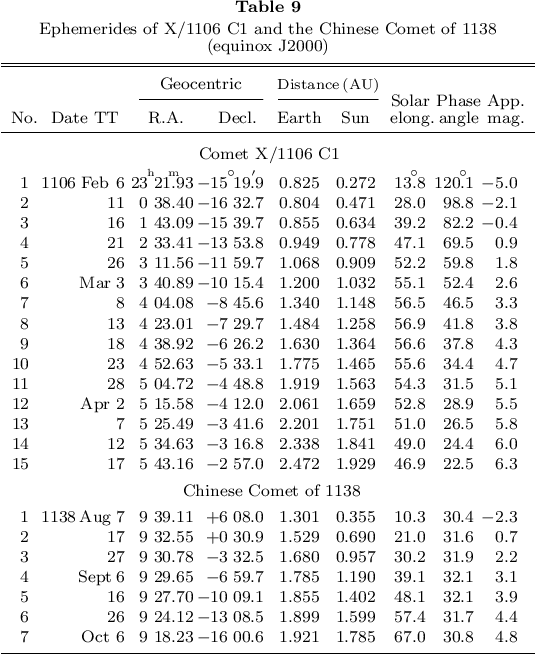}}}
\vspace{0cm}
\end{table}

\section{Concluding Remarks} 
The purpose of this paper has been to investigate the generation
of Kreutz sungrazers immediately prior to the current one, that
is, the direct parents.   In the context of the contact-binary
model (Sekanina 2021), there may exist a number of sungrazers of
that generation (reaching perihelion between, say, the 10th and
13th centuries), but the foremost among them are the members of
Populations~I and II, the Great Comet of 1106 and the Chinese
comet of 1138 (No.~403 in Ho's catalogue), respectively.  The
contact-binary model was initially based on a postulate that the
Great March Comet of 1843 was the largest surviving mass of the 1106
comet, whereas the Great September Comet of 1882 was the largest
surviving mass of another comet, because at the time the model was
conceived (late summer 2021), the 1106 comet still was the only feasible
candidate for a Kreutz sungrazer of its generation.  More recently,
the problem of a missing second sungrazer was solved, resulting in
the ``discovery'' of the 1138 comet and offering a virtual proof,
by orbit integration, of its feasibility as a member of Population~II
of its generation and the parent to the 1882 sungrazer and Ikeya-Seki
(Sekanina \& Kracht 2022).

As the best estimate for its perihelion time is August~1, the
1138 comet can be used to examine the standard narrative that
(in the era prior to Sun-exploring spacecraft) a Kreutz sungrazer
reaching perihelion at the ``wrong'' time of the year\,---\,between
mid-May and mid-August\,--- was missed {\it if not seen in
daylight\/}.  My review of this scenario, with the comet
approaching and leaving perihelion from behind the Sun, has led
to a conclusion that the geometry involved dictated that it
was at close proximity to perihelion --- in daylight --- when
the singularly unfavorable observing conditions took place.
This included a protracted period of extremely small solar
elongations, as well as a very short dust tail, hook-shaped and
turning sunward.  On the other hand, as the comet receded a
little beyond 1~AU from the Sun, its solar elongation increased
enough and its fading was modest enough, that over a period of
at most a few weeks it reached, in astronomical twilight before
sunrise, elevations near 10$^\circ$, when sighted from a moderate
northern latitude.   Its dust tail, still hook shaped, stretched
antisunward to elevations of $\sim$15--20$^\circ$, and the plasma
tail may have extended up to $\sim$30$^\circ$ above the horizon.
Accordingly, the notion that sungrazers with perihelia at the
``wrong'' time of the year were missed {\it unless seen in
daylight\/} is misleading.

This was an issue with other sungrazers or potential sungrazers as well,
such as the controversial comet(s) of 1041 and, perhaps surprisingly,
comet Pereyra of 1963, whose perihelion passage was just outside the
zone of ``wrong'' time.  Even though intrinsically brighter than
Ikeya-Seki, Pereyra was undetected in daylight and discovered only
about three weeks after perihelion.  The comet of 1138 thus was no
exception.

The mediocrity of the daylight observing conditions for the 1138 comet is
particularly well illustrated in juxtaposition with the greatly superior
conditions for the Great Comet of 1106.  This is not surprising, as the
Earth was at almost exactly opposite points of its orbit at the
times the two objects reached perihelion.  Comparison has been carried
out by applying a criterion based on Schaefer's (1993, 1998) naked-eye
limiting-magnitude model, leading to a conclusion that the vast
differences were triggered by three causes:\ a disparity in geocentric
distance; a related disparity in solar elongation, the 1138 comet always
projecting closer to the Sun than the 1106 comet at the same heliocentric
distance; and a disparity in phase correction, namely strong forward
scattering by microscopic dust early after the 1106 comet's
perihelion.{\vspace{0.04cm}}\,

In the light of these findings, the account by Sigebert de Gembloux
of what turned out to be the daylight discovery of the 1106 comet
on February~2 has been most welcome.  The detection is not only in
line with the predicted generally favorable observing conditions for
this object, but even more so in the earliest post-perihelion times.
The comet had apparently passed perihelion only hours before the
reported observation was made.{\vspace{0.03cm}}

Next I have addressed the relationship between the last glimpse of a
historical sungrazer (i.e., its tail) and a predicted apparent magnitude
of its head at the time.  Hasegawa \& Nakano (2001) assumed magnitude~5,
but most of their Kreutz candidates were intrinsically much too bright.
Based on reports for recent sungrazers, I concluded that the
relationship followed a rule of thumb, with the tail still marginally
visible when the head was long below the naked-eye detection limit and
telescopically estimated at magnitude $\sim$7 (Sekanina 2022).  However,
all sungrazers involved in the experiment were those with highly favorable
observing conditions.  On the other hand, I have now found that the
disappearance of the 1138 comet coincided with the head's predicted apparent
magnitude of 4.5, or 10 times brighter, in agreement with the result
for the September comet of 1041.  It appears that the apparent-magnitude
threshold in the rule of thumb correlates with the observing conditions,
an issue that is still waiting to be resolved.{\vspace{0.04cm}}

The final addressed subject has been the appearance and morphology of
the nucleus after its tidal fragmentation at perihelion.  From limited
evidence of such events at the Sun and Jupiter, it is not unusual to see
the fragmented nucleus consisting of a number of nearly equidistant
condensations buried in an oval cloud of debris, with the brightest ones
located generally near the middle of the chain.  For the comet of 1106,
nuclear fragmentation has been almost a nonissue, in part because there
is no evidence for splitting of the 1843 sungrazer.  To my knowledge,
there is only one other comet --- C/1668~E1 --- that could have derived
from the 1106 comet, making the 1843 sungrazer the principal fragment.
A vast majority of the dwarf sungrazers, detected by coronagraphs on
board the Solar and Heliospheric Observatory and other Sun-exploring
spacecraft, appears to be minor fragments of the 1106 comet, but that
is another matter.  More to the point is the possibility that additional
fragments could arrive later in the 21st century and beyond.{\vspace{0.04cm}}

Of greater interest is the comet of 1138, because two of its components
were observed:\ the 1882 sungrazer (the principal fragment) and Ikeya-Seki.
In addition, there are two earlier suspects --- X/1702~D1 and the sun-comet
of 1792 --- and a predicted candidate, expected to arrive within 10~years
or so from now.  The fragmented nucleus of the 1138 comet is thus modeled
on the assumption that it consisted of five condensations.  Its length at
the time of the comet's recorded tail sighting, $\sim$33~days after 
perihelion, is calculated to have amounted to about
15$^{\prime\prime}\!$.\,\,\,\,

To learn about changes in the overall appearance and morphology that a
fragmented nucleus undergoes over a period of time, I present a selected
set of four drawings of the nuclear region of the 1882 sungrazer, made by
W.~L.~Elkin at the Cape Observatory between October~13 and November~21,
25 to 64~days after perihelion.  All four condensations, for which the
positions were calculated from Kreutz's orbits, are seen only in the
first drawing.  The separation distances measured by Elkin are in very
good agreement with the computed ones.

In the other three drawings fewer than four condensations are apparent,
most of them marginally, with the principal fragment~B obscured by
the cloud of debris to the extent that it is completely invisible
(with one possible exception).  Two drawings are dominated by fragment~A,
which during October was evidently undergoing a protracted flare-up.
By the end of the month it was already fading and nearly gone by
November~21.

The 1882 sungrazer was under observation until the end of May 1883, and
unlike fragments B and/or C, fragment~A was then no longer seen.  Its
persisting flare-up in October was the kind of event that is often
exhibited by short-lived companions of split comets and potentially is
a signature of a disintegrating object.  It needs to be admitted that
there is no correlation between the mass and instantaneous brightness
of a sungrazer's fragment and, as a corollary, that not every bright
enough fragment, which is given birth in the process of tidal
fragmentation, does necessarily survive throughout the orbit about
the Sun to successfully pass --- as an essentially intact, sizable
sungrazer --- yet another perihelion.

This caveat applies to the 1138 comet as much as to the 1882 sungrazer,
with one exception:\ fragments of these two objects cannot readily be
compared because of the difference of one generation between them.
Because the fragments of the 1882 sungrazer are fragments of a fragment
of the 1138 comet, survival throughout the revolution about the Sun
is a priori more likely for the fragments of the 1138 comet than the
1882 comet.  Still, survival is not warranted for either of them.
This uncertainty is a major complication for predictions of future
appearances of Kreutz sungrazers that is in principle impossible
to bypass.

Concluding this investigation, it is my hope that~it~has cleared
any persisting ambiguity or confusion regarding the parents of
the current generation of Populations~I and II of the Kreutz
system, and once and for all put to rest the lingering problem of
a missing second sungrazer.

\pagebreak
\begin{center} 
{\footnotesize REFERENCES}
\end{center}
\vspace{-0.45cm}
\begin{description}
{\footnotesize
\item[\hspace{-0.3cm}]
Bennett, J.\ C., \& Venter, S.\ C.\ 1966, Mon.\ Not.\ Astron.\ Soc.\ South{\linebreak}
 {\hspace*{-0.6cm}}Africa, 25, 12
 \\[-0.57cm]
\item[\hspace{-0.3cm}]
Bennett, J.\ C., van Zyl, L.\ L., \& Venter, S.\ C.\ 1964, Mon.\ Not.{\linebreak}
 {\hspace*{-0.6cm}}Astron.\ Soc.\ South Africa, 23, 45
\\[-0.57cm]
%
%
\item[\hspace{-0.3cm}]
Capen, C.\ F.\ 1964, Stroll.\ Astron., 17, 178
\\[-0.57cm]
\item[\hspace{-0.3cm}]
England, K.\ J.\ 2002, J.\ Brit.\ Astron.\ Assoc., 112, 13
\\[-0.57cm]
%
%
\item[\hspace{-0.3cm}]
Gill, D., Finlay, W.\ H., \& Elkin, W.\ L.\ 1911, Ann.\ Cape Obs., 2{\linebreak}
 {\hspace*{-0.6cm}}(Part 1), 3
\\[-0.57cm]
%
%
\item[\hspace{-0.3cm}]
Hagughney, L.\ C., Bader, M., \& Innes, R.\ 1967, Astron.\ J., 72, 1166
\\[-0.57cm]
\item[\hspace{-0.3cm}]
Hasegawa, I.\ 1979, Publ.\ Astron.\ Soc.\ Japan, 31, 257
\\[-0.57cm]
\item[\hspace{-0.3cm}]
Hasegawa, I., \& Nakano, S.\ 2001, Publ.\ Astron.\ Soc.\ Japan, 53, 931
\\[-0.57cm]
%
%
\item[\hspace{-0.3cm}]
Ho, P.-Y.\ 1962, Vistas Astron., 5, 127
\\[-0.57cm]
\item[\hspace{-0.3cm}]
Hubbard, J.\ S.\ 1851, Astron.\ J., 2, 57
\\[-0.57cm]
\item[\hspace{-0.3cm}]
Hubbard, J.\ S.\ 1852, Astron.\ J., 2, 153
\\[-0.57cm]
\item[\hspace{-0.3cm}]
Hufnagel, L.\ 1919, Astron.\ Nachr., 209, 17
\\[-0.57cm]
%
%
\item[\hspace{-0.3cm}]
Kreutz, H.\ 1888, Publ.\ Sternw.\ Kiel, No.\ 3
\\[-0.57cm]
\item[\hspace{-0.3cm}]
Kreutz, H.\ 1891, Publ.\ Sternw.\ Kiel, No.\ 6
\\[-0.57cm]
\item[\hspace{-0.3cm}]
Kreutz, H.\ 1901, Astron.\ Abhandl., 1, 1
\\[-0.57cm]
\item[\hspace{-0.3cm}]
Krishan, V,, \& Sivaraman, K.\ R.\ 1982, Moon Plan., 26, 209
\\[-0.57cm]
\item[\hspace{-0.3cm}]
Krishna Swamy, K.\ S.\ 1978, Astrophys.\ Space Sci., 57, 491
\\[-0.57cm]
\item[\hspace{-0.3cm}]
Kronk, G.\ W.\ 1999, Cometography, Volume 1:\ Ancient--1799.{\linebreak}
 {\hspace*{-0.6cm}}Cambridge, UK:\ University Press, 580pp
\\[-0.57cm]
%
%
\item[\hspace{-0.3cm}]
Marcus, J.\ N.\ 2007, Int.\  Comet Quart., 29, 39
\\[-0.57cm]
\item[\hspace{-0.3cm}]
Marsden, B.\ G.\ 1967, Astron.\ J., 72, 1170
\\[-0.57cm]
%
%
\item[\hspace{-0.3cm}]
Marsden, B.\ G., \& Williams, G.\ V.\ 2008, Catalogue of Cometary{\linebreak}
 {\hspace*{-0.6cm}}Orbits 2008, 17th ed.  Cambridge, Mass.:\ Minor Planet
 Center/{\linebreak}
 {\hspace*{-0.6cm}}Central Bureau for Astronomical Telegrams, 195pp
\\[-0.57cm]
\item[\hspace{-0.3cm}]
Mart\'{\i}nez Us\'o, M.\ J., \& Marco Castillo, F.\ J.\ 2023, in
 Asteroids,{\linebreak}
 {\hspace*{-0.6cm}}Comets, Meteors 2023, Lunar Plan.\ Inst.\ Contr.\ 2851,
 id.~2082
\\[-0.57cm]
\item[\hspace{-0.3cm}]
Matyagin, V.\ S., Sabitov, S.\ N., \& Kharitonov, A.\ V.\ 1968, Sov.{\linebreak}
 {\hspace*{-0.6cm}}Astron., 11, 863
\\[-0.57cm]
\item[\hspace{-0.3cm}]
Schaefer, B.\ E.\ 1993, Vistas Astron., 36, 311
\\[-0.57cm]
\item[\hspace{-0.3cm}]
Schaefer, B.\ E.\ 1998, Sky Tel., 95, 57; code version by L.\ Bogan{\linebreak}
 {\hspace*{-0.6cm}}at {\tt https://www.bogan.ca/astro/optics/vislimit.html}
\\[-0.57cm]
\item[\hspace{-0.3cm}]
Schwab, F.\ 1883, Astron.\ Nachr., 105, 3
\\[-0.57cm]
%
%
\item[\hspace{-0.3cm}]
Sekanina, Z.\ 2000, Astrophys.\ J., 542, L147
\\[-0.57cm]
\item[\hspace{-0.3cm}]
Sekanina, Z.\ 2002, Astrophys.\ J., 566, 577
\\[-0.57cm]
\item[\hspace{-0.3cm}]
Sekanina, Z.\ 2021, eprint arXiv:2109.01297
\\[-0.57cm]
%
%
%
\item[\hspace{-0.3cm}]
Sekanina, Z.\ 2022, eprint arXiv:2202.01164
\\[-0.57cm]
%
%
\item[\hspace{-0.3cm}]
Sekanina, Z.\ 2023, eprint arXiv:2310.05320
\\[-0.57cm]
\item[\hspace{-0.3cm}]
Sekanina, Z.\ 2024, eprint arXiv:2404.00887
\\[-0.57cm]
%
%
\item[\hspace{-0.3cm}]
Sekanina, Z.\ 2025, eprint arXiv:2503.15467
\\[-0.57cm]
%
%
%
%
%
%
\item[\hspace{-0.3cm}]
Sekanina, Z., \& Kracht, R.\ 2022, eprint arXiv:2206.10827
\\[-0.57cm]
\item[\hspace{-0.3cm}]
Shanklin, J.\ 2004, J.\,Brit.\,Astron.\,Assoc.,\,114,\,44; also:\ The Comet's{\linebreak}
 {\hspace*{-0.6cm}}Tale,\,10,\,\#2,\,4 (October 2003); online:\
 {\tt https:/$\!$/proama\&g.pdf}
\\[-0.57cm]
\item[\hspace{-0.3cm}]
Steele, J.\ M.\ 2008, A Brief Introduction to Astronomy~in~the~Mid-{\linebreak}
 {\hspace*{-0.6cm}}dle East.  London:\ Saqi Books Publ., 140pp
\\[-0.57cm]
\item[\hspace{-0.3cm}]
Stephenson, F.\ R., \& Fatoohi, L.\ J.\ 1994, J.\ Hist.\ Astron., 25, 99
\\[-0.57cm]
\item[\hspace{-0.3cm}]
Stone, M.\ H.\ 2014, J.\ Anthrop., 2014, 1 
\\[-0.57cm]
\item[\hspace{-0.3cm}]
Strom, R.\ 2002, Astron.\ Astrophys., 387, L17
\\[-0.57cm]
\item[\hspace{-0.3cm}]
Thompson, W.\ T.\ 2009, Icarus, 200, 351
\\[-0.57cm]
\item[\hspace{-0.3cm}]
Thompson, W.\ T.\ 2015, Icarus, 261, 122
\\[-0.57cm]
%
%
\item[\hspace{-0.3cm}]
Warner, B.\ 1980, Mon.\ Not.\ Astron.\ Soc.\ South Africa, 39, 69
\\[-0.57cm]
\item[\hspace{-0.3cm}]
Weinberg, J.\ L., \& Beeson, D.\ E.\ 1976a, in The Study of Comets,{\linebreak}
 {\hspace*{-0.6cm}}NASA\,SP-393,\,ed.$\:$B.\,Donn,\,M.\,Mumma,\,W.\,Jackson,\,M.\,A'Hearn,{\linebreak}
 {\hspace*{-0.6cm}}\& R.\ Harrington. Washington, DC:\ U.S.\ GPO, 92
\\[-0.65cm]
\item[\hspace{-0.3cm}]
Weinberg, J.\ L., \& Beeson, D.\ E.\ 1976b, Astron.\ Astrophys.,~48,~151}
%
%
\vspace{-0.72cm}
\end{description}

\end{document}